\documentclass[oneside,11pt]{article}

\usepackage[utf8]{inputenc}
\usepackage[T1]{fontenc}

\usepackage{amsthm}
\usepackage{amsmath}
\usepackage{amsfonts}
\usepackage{amssymb}

\usepackage{graphicx}
\usepackage{float}
\usepackage{booktabs}
\usepackage{multirow}
\usepackage{rotating}
\usepackage{array}

\usepackage{nowidow}

\usepackage{breakcites}

\usepackage{enumitem}

\usepackage[top=1.5cm,bottom=2cm,right=2.5cm,left=2.5cm]{geometry}
\usepackage{titling}
\usepackage{natbib}

\usepackage{ifplatform}

\ifwindows
\fi

\DeclareFontFamily{OT1}{pzc}{}
\DeclareFontShape{OT1}{pzc}{m}{it}{<-> s * [1.10] pzcmi7t}{}
\DeclareMathAlphabet{\mathpzc}{OT1}{pzc}{m}{it}

\usepackage[pdftex,final,bookmarksnumbered,bookmarksopen=false,breaklinks,colorlinks]{hyperref}
\hypersetup{
	final,
    bookmarksnumbered=true,
    bookmarksopen=true,
    bookmarksopenlevel=0,
    unicode=false,          
    pdftoolbar=true,        
    pdfmenubar=true,        
    pdffitwindow=false,     
    pdftitle={A generative angular model of protein structure evolution},    
	pdfdisplaydoctitle=true, 
	pdftoolbar=true,
	pdfmenubar=true,
	pdflang={English},
    pdfauthor={Michael Golden, Eduardo Garcia-Portugues, Michael Sorensen, Kanti V. Mardia, Thomas Hamelryck, Jotun Hein},     
    pdfsubject={arXiv paper},   
    pdfcreator={Eduardo Garcia-Portugues},   
    pdfproducer={Eduardo Garcia-Portugues}, 
    pdfkeywords={Evolution}{Protein structure}{Probabilistic model}{Directional statistics}, 
    pdfnewwindow=true,      
    breaklinks=true,
    hidelinks,
    linkcolor=black,          
    citecolor=black,        
    filecolor=magenta,      
    urlcolor=cyan           
}

\allowdisplaybreaks

\newif\iffigures
\figurestrue

\newif\ifmain
\maintrue
\newif\ifsupplement
\supplementtrue

\begin{document}

\ifmain

\title{A generative angular model of protein structure evolution}
\setlength{\droptitle}{-1cm}
\predate{}%
\postdate{}%

\date{}

\author{
Michael Golden$^{1,6}$, Eduardo Garc\'ia-Portugu\'es$^{2,3}$, Michael S\o rensen$^{3}$,\\ Kanti V.~Mardia$^{4}$, Thomas Hamelryck$^{2,5}$, and Jotun Hein$^{1}$}

\footnotetext[1]{
Department of Statistics, University of Oxford (UK).}
\footnotetext[2]{
Bioinformatics Centre, Section for Computational and RNA Biology, Department of Biology, University of Copenhagen (Denmark).}
\footnotetext[3]{
Department of Mathematical Sciences, University of Copenhagen (Denmark).}
\footnotetext[4]{
Department of Mathematics, University of Leeds (UK).}
\footnotetext[5]{
Image Section, Department of Computer Science, University of Copenhagen (Denmark).}
\footnotetext[6]{Corresponding author. e-mail: \href{mailto:golden@stats.ox.ac.uk}{golden@stats.ox.ac.uk}.}
\maketitle

\begin{abstract}
Recently described stochastic models of protein evolution have demonstrated that the inclusion of structural information in addition to amino acid sequences leads to a more reliable estimation of evolutionary parameters. We present a generative, evolutionary model of protein structure and sequence that is valid on a local length scale. The model concerns the local dependencies between sequence and structure evolution in a pair of homologous proteins. The evolutionary trajectory between the two structures in the protein pair is treated as a random walk in dihedral angle space, which is modelled using a novel angular diffusion process on the two-dimensional torus. Coupling sequence and structure evolution in our model allows for modelling both ``smooth'' conformational changes and ``catastrophic'' conformational jumps, conditioned on the amino acid changes. The model has interpretable parameters and is comparatively more realistic than previous stochastic models, providing new insights into the relationship between sequence and structure evolution. For example, using the trained model we were able to identify an apparent sequence-structure evolutionary motif present in a large number of homologous protein pairs. The generative nature of our model enables us to evaluate its validity and its ability to simulate aspects of protein evolution conditioned on an amino acid sequence, a related amino acid sequence, a related structure or any combination thereof.
\end{abstract}
\begin{flushleft}
\end{flushleft}

\section{Introduction}
\label{sec:Intro}

Recently, several studies \citep{challis2012stochastic, herman2014simultaneous} have proposed joint stochastic models of evolution which take into account simultaneous alignment of protein sequence and structure. These studies point out the limitations of earlier non-probabilistic methods, which often rely on heuristic procedures to infer parameters of interest. A major disadvantage of using heuristic procedures is that they typically fail to account for sources of uncertainty. For example, they may rely on a single fixed alignment of proteins, which is highly unlikely to be the \textit{true} underlying alignment. Such a fixed alignment may bias the inference of the distribution over evolutionary\nolinebreak[4] trees.\\

We present a generative evolutionary model, ETDBN (Evolutionary Torus Dynamic Bayesian Network) for pairs of homologous proteins. ETDBN captures dependencies between sequence and structure evolution, accounts for alignment uncertainty, and models the local dependencies between aligned sites. This extends the models presented in \citet{challis2012stochastic} and \citet{herman2014simultaneous}, which emphasise estimation of evolutionary parameters such as the evolutionary time between species, tree topologies and alignment.\\

ETDBN is motivated by the non-evolutionary TorusDBN model \citep{boomsma2008generative}. TorusDBN is a first-order Hidden Markov Model (HMM) that represents a single protein structure as a sequence of $\phi,\psi$ dihedral angle pairs, which are modelled using continuous bivariate angular distributions \citep{frellsen2012towards}. Likewise, ETDBN treats protein structure evolution as a random walk in space, again making use of the $\phi$ and $\psi$ dihedral angles (top of Figure~\ref{fig:hmm_diagram}).

\begin{figure}[h!]
\iffigures
\centering
\includegraphics[width=0.8\textwidth]{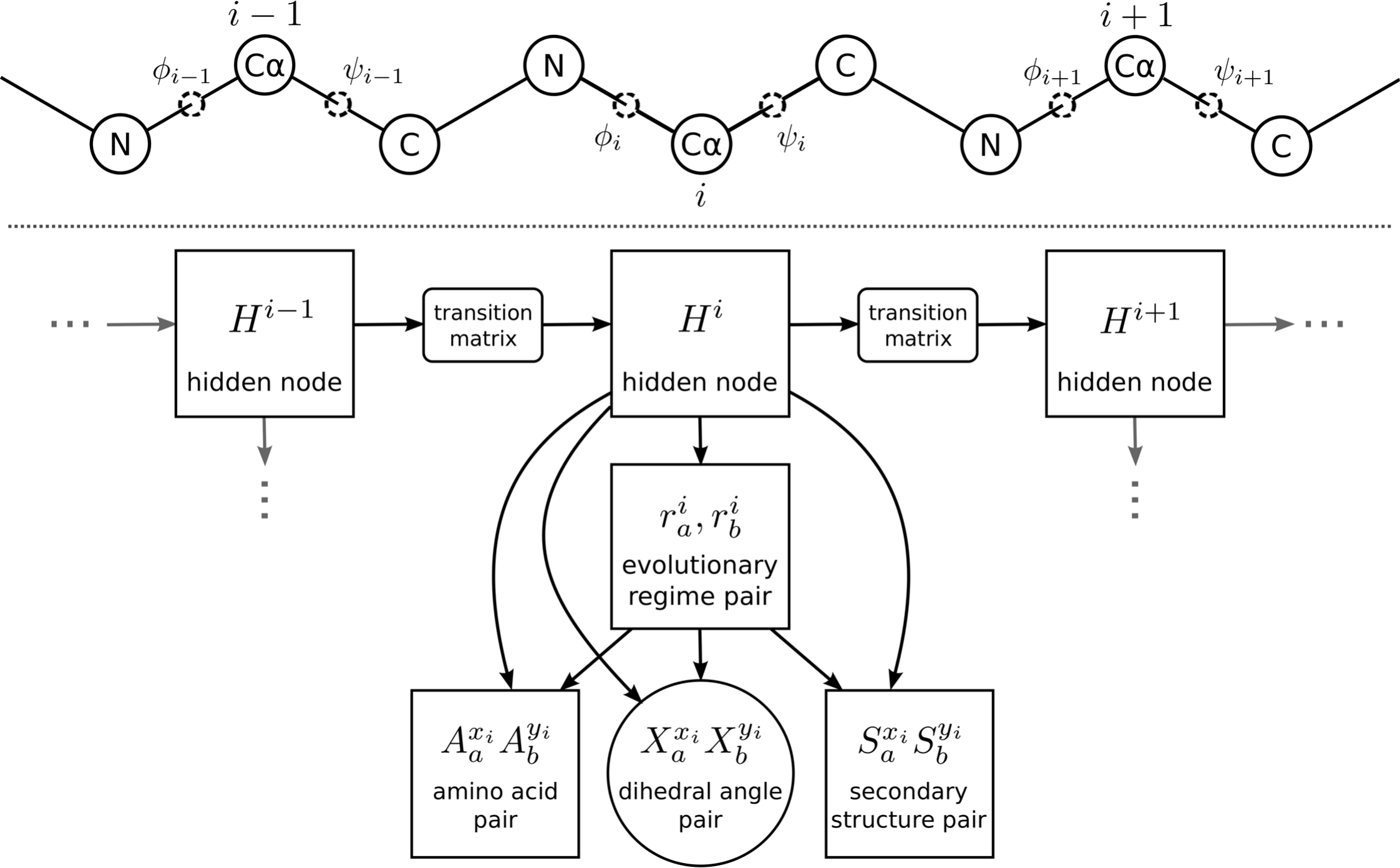}
\fi
\caption{\small Above: dihedral angle representation. A small section of a single protein backbone (three amino acids) with $\phi$ and $\psi$ dihedral angles shown, together with $C\alpha$ atoms which attach to the amino acid side-chains. Each amino acid side-chain determines the characteristic nature of each amino acid. Every amino acid position corresponds to a hidden node in the HMM below. Note that we only show a single protein, whereas the model considers a pair. Below: depiction of HMM architecture of ETDBN where each $H$ along the horizontal axis represents an evolutionary hidden node. The horizontal edges between evolutionary hidden nodes encode neighbouring dependencies between aligned sites. The arrows between the evolutionary hidden nodes and evolutionary regime pair nodes encode the conditional independence between the observation pair variables $A_a^{x_i},A_b^{y_i}$ (amino acid site pair), $X_a^{x_i},X_b^{y_i}=\langle\phi_a^{x_i},\psi_a^{x_i}\rangle,\langle\phi_b^{y_i},\psi_b^{y_i}\rangle$ (dihedral angle site pair) and $S_a^{x_i},S_b^{y_i}$ (secondary structure class site pair) The circles represent continuous variables and the rectangles represent discrete variables.}
  \label{fig:hmm_diagram}%
\end{figure}

Each amino acid in a peptide chain is covalently bonded to the next via a chemical bond referred to as a \textit{peptide bond}. Peptide bonds have a partial double bond nature, that results in a planar configuration of atoms in space. This configuration allows the protein backbone structure to be largely described in terms of a series of $\phi$ and $\psi$ dihedral angles that defines the relationship between the planes in three-dimensional space. A benefit of this representation is that it bypasses the need for structural alignment, unlike in \citet{challis2012stochastic} and \citet{herman2014simultaneous}, which is required to simultaneously superimpose structures in three-dimensional space for comparison purposes. Furthermore, dihedral angles provide a natural character state for aligning positions in the absence of amino acid characters.\\

The evolution of dihedral angles in ETDBN is modelled using a novel stochastic diffusion process developed in \citet{garciap2017diffusions}. In addition to this, a coupling is introduced such that an amino acid change can lead to a \textit{jump} in dihedral angles and a change in diffusion process, allowing us to capture changes in amino acid that are coupled with changes in dihedral angle or secondary structure. As in \citet{challis2012stochastic} and \citet{herman2014simultaneous}, the insertion and deletion (indel) evolutionary process is also modelled in order to account for alignment uncertainty \citep{thorne1992inching}.\\

Given that naturally occurring protein sequences are evolutionarily related, it is only natural to consider models that take such evolution into account. Two or more homologous proteins will share a common ancestor, which leads to underlying tree-like dependencies. These dependencies manifest themselves most noticeably in the degree of amino acid sequence similarity between two homologous proteins. The strength of these dependencies is assumed to be a result of two main factors: the time since the common ancestor and the rate of evolution. If the rate of evolution is assumed constant, recently separated proteins will typically be similar in both sequence and structure, whereas proteins which share a common ancestor in the distant past will plausibly appear independent of one another.\\

Failing to account for evolutionary dependencies can lead to false conclusions \citep{felsenstein1985phylogenies}, whereas accounting for evolutionary dependencies allows information from homologous proteins to be incorporated in a principled manner. This can lead to more accurate inferences, such as the prediction of a protein structure from a homologous protein sequence and structure, known as homology modelling. An example of a popular homology modelling software environment is SWISS-MODEL \citep{arnold2006swiss}. Stochastic models such as ETDBN are not expected to compete with packages such as SWISS-MODEL. However, they allow for estimation of evolutionary parameters and statements about uncertainty to be made in a statistically rigorous manner.\\

Parameters of ETDBN, particularly those governing the amino acid and dihedral angle evolutionary processes, were learnt during a training phase using 1200 homologous protein pairs from the HOMSTRAD database \citep{mizuguchi1998homstrad}. This resulted in a realistic prior distribution over proteins compared to previous stochastic models, enabling biological insights into the relationship between sequence and structure evolution, such as patterns of amino acid change that are informative of patterns of structural change \citep{grishin2001fold}. It was with these features in mind that ETDBN was developed.

\section{Evolutionary model}
\label{sec:Model}

\subsection{Overview}

ETDBN is a dynamic Bayesian network model of local protein sequence and structure evolution along a pair of aligned homologous proteins $p_{a}$ and $p_{b}$. ETDBN can be can be viewed as an HMM (see Figure~\ref{fig:hmm_diagram}). Each hidden node of the HMM, corresponding to an aligned position, adopts an evolutionary hidden state specifying a distribution over three different observations pairs: a pair of amino acid characters, a pair of dihedral angles and a pair of secondary structures classifications. A transition probability matrix specifies neighbouring dependencies between adjacent evolutionary states. For example, it is expected that evolutionary hidden states specifying predominantly $\alpha$-helical secondary structure configurations will occur adjacent to one another. Finally, each hidden state specifies a distribution over a pair of evolutionary regimes at each aligned position. This introduces a stronger coupling between changes in amino acid that are informative of changes in dihedral angle or secondary structure between proteins, whilst simultaneously enabling the possibility of a \textit{jump} -- a large change in dihedral angle or secondary structure, (e.g. helix to sheet) coupled with specific amino acid changes at an aligned position.

\subsection{Observation types}

The two proteins, $p_a$ and $p_b$, in a homologous pair are associated with a pair of observation sequences $O_{a}$ and $O_{b}$ obtained from experimental data, respectively. An $i$th site observation pair, $O^{i}=(O_{a}^{x(i)},O_{b}^{y(i)})$, is associated with every aligned site $i$ in an alignment $M_{ab}$ of $p_{a}$ and $p_{b}$, where $M_{ab}^{i}\in\left\{ \left(\substack{x\\
y
}
\right),\left(\substack{x\\
-
}
\right),\left(\substack{-\\
y
}
\right)\right\}$  specifies the homology relationship at position $i$ of the alignment (homologous, deletion with respect to $p_a$ and insertion with respect to $p_a$, respectively.), $i$ is taken to run from 1 to $m$, $m$ is the length of the alignment $M_{ab}$, and $x\in\{1,\ldots,|p_{a}|\}$ and $y\in\{1,\ldots,|p_{b}|\}$ specify the indices of the positions in $p_a$ and $p_b$, respectively. $|p_{a}|$ and $|p_{a}|$ specify the number of sites in $p_a$ and $p_b$, respectively.\\

Each site observation, $O_{a}^{x(i)}$ and $O_{b}^{y(i)}$, contains amino acid and structural information corresponding to the two $C\alpha$ atoms at aligned site $i$ belonging to each of the two proteins. A site observation corresponding to a particular protein at aligned site $i$, $O_{a}^{x(i)}$, is comprised of three different data types associated with the $C\alpha$  atom: an amino acid ($A_{a}^{x(i)}$; discrete, one of twenty canonical amino acids), $\phi$ and $\psi$ dihedral angles ($X_{a}^{x(i)}=\langle\phi_{a}^{x(i)},\psi_{a}^{x(i)}\rangle$; continuous, bivariate), and a secondary structure classification ($S_{a}^{x(i)}$; discrete, one of three classes: helix (H), sheet (S) or coil (C)). Therefore, $O_{a}^{x(i)}=(A_{a}^{x(i)},X_{a}^{x(i)},S_{a}^{x(i)})$ and $O_{b}^{y(i)}=(A_{b}^{y(i)},X_{b}^{y(i)},S_{b}^{y(i)})$.

\subsection{Model structure}

The sequence of hidden nodes in the HMM is written as $H=(H^{1},H^{2},\ldots,H^{m}$). Each hidden node $H^{i}$ in the HMM corresponds to a site observation pair, $O_{a}^{x(i)}$ and $O_{b}^{y(i)}$, at an aligned site $i$ in the alignment $M_{ab}$. Initially we treat the alignment $M_{ab}$ as given \textit{a priori}, but later modify the HMM to marginalise out an unobserved alignment.\\

The model is parametrised by $h$  hidden states. Every hidden node $H^{i}$ corresponding to an aligned site $i$ takes an integer value from $1$  to $h$ for the hidden state at node $H^{i}$. In turn, each hidden state specifies a distribution over an evolutionary regime pair: $(r_{a}^{i},r_{b}^{i})$ as a function of evolutionary time. An evolutionary regime pair consists of two evolutionary regimes:  $r_{a}^{i}$ and $r_{b}^{i}$. Each of the two evolutionary regimes takes an integer value 1 or 2, \textit{i.e.} $(r_{a}^{i},r_{b}^{i})\in \{(1,1),(1,2),(2,1),(2,2)\}$. We return to the specific role of the evolutionary regimes pairs in the next section.\\

The state of $H^i$ together with the evolutionary regime pair, $(r_{a}^{i},r_{b}^{i})$,  and the evolutionary time separating proteins $p_a$ and $p_b$, $t_{ab}$, specify a distribution over three conditionally independent stochastic processes describing each of the three types of site observation pairs: $A^{i}=(A_{a}^{x(i)},A_{b}^{y(i)})$, $X^{i}=(X_{a}^{x(i)},X_{b}^{y(i)})$  and $S^{i}=(S_{a}^{x(i)},S_{b}^{y(i)})$. This conditional independence structure allows the likelihood of a site observation pair at an aligned site $i$ to be written as follows:
\begin{align}
p(O^{i}|H^{i},r_{a}^{i},r_{b}^{i},t_{ab}) = \overset{\text{amino acid evolution}}{\overbrace{p(A^{i}|H^{i},r_{a}^{i},r_{b}^{i},t_{ab})}} \times \overset{\text{dihedral angle evolution}}{\overbrace{p(X^{i}|H^{i},r_{a}^{i},r_{b}^{i},t_{ab})}}
\,\times \!\!\!\overset{\text{secondary structure evolution}}{\overbrace{p(S^{i}|H^{i},r_{a}^{i},r_{b}^{i},t_{ab})}}.
\label{eq:conditional_independence}
\end{align}
The assumption of conditional independence provides computational tractability, allowing us to avoid costly marginalisation when certain combinations of data are missing (e.g. amino acid sequences present, but secondary structures and dihedral angles missing).

\subsection{Stochastic processes: modelling evolutionary dependencies}

Each evolutionary regime couples together three time-reversible stochastic processes that separately describe the evolution of the three pairs of observation types, as in equation (\ref{eq:conditional_independence}). Each evolutionary regime is intended to capture different features of sequence and structural evolution. Parameters that correspond to a particular evolutionary regime are termed \textit{regime-specific}, whereas parameters that are shared across all evolutionary regimes are termed \textit{global}.

\paragraph{Amino acid evolution}
As is typical with models of sequence evolution, amino acid evolution, $p(A_{a}^{x(i)},A_{b}^{y(i)}|H^{i},r_{a}^{i},r_{b}^{i},t_{ab})$, is described by a Continuous-Time Markov Chain (CTMC). Each amino acid CTMC is parametrised in the following way: the exchangeability of amino acids is described by a $20\times20$ symmetric global exchangeability matrix $S$ (190 free parameters; \citet{whelan2001general}), a regime-specific set of 20 amino acid equilibrium frequencies $\Pi_{r}^{h}=\text{diag}\{\pi_{1},\pi_{2},\ldots,\pi_{20}\}$ (19 free parameters per evolutionary regime) and a regime-specific scaling factor $\Lambda_{r}^{h}$ (1 free parameter per evolutionary regime). Together these parameters define a regime-specific time-reversible amino acid rate matrix $Q_{r}^{h}=\Lambda_{r}^{h}S\Pi_{r}^{h}$. The stationary distribution of $Q_{r}^{h}$ is given by the amino acid equilibrium frequencies: $\Pi_{r}^{h}$.

\paragraph{Secondary structure evolution}
Secondary structure evolution, $p(S_{a}^{x(i)},S_{b}^{y(i)}|H^{i},r_{a}^{i},r_{b}^{i},t_{ab})$, is also described by a CTMC. In our model we use three discrete classes to describe secondary structure at each position: helix (H), sheet (S) and random coil (C).\\

The exchangeability of secondary structure classes at a position is described by a $3\times3$  symmetric global exchangeability matrix $V$ and a regime-specific set of 3 secondary structure equilibrium frequencies $\Omega_{r}^{h}=\text{diag}\{\pi_{1},\pi_{2},\pi_{3}\}$. Together they define a regime-specific time-reversible secondary structure rate matrix $R_{r}^{h}=V\Omega_{r}^{h}$, with stationary distribution: $\Omega_{r}^{h}$.

\paragraph{Dihedral angle evolution}
Central to our model is evolutionary dependence between dihedral angles, $p(X_{a}^{x(i)},X_{b}^{y(i)}|H^{i},r_{a}^{i},r_{b}^{i},t_{ab})$. Typically, the continuous-time evolution of the continuous-state random variables is modelled by a diffusive process such as the Ornstein--Uhlenbeck (OU) process, as in \citet{challis2012stochastic}. However, an OU process is not appropriate for dihedral angles as they have a natural periodicity. For this reason, a bivariate diffusion that captures the periodic nature of dihedral angles, the \textit{Wrapped Normal} (WN) \textit{diffusion}, was specifically developed for this paper in \citet{garciap2017diffusions}.\\

\begin{figure}[h!]
\iffigures
\centering
\includegraphics[width=0.55\textwidth]{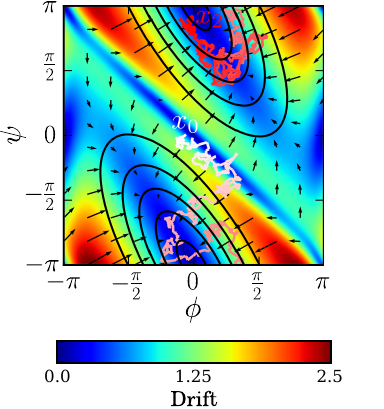}
\fi
\caption{\small Drift vector field for the WN diffusion with $A=(1, 0.5; 0.5, 0.5)$, $\mu=(0,0)$ and $\Sigma=(1.5)^2I$. The colour gradient represents the Euclidean norm of the drift. The contour lines represent the stationary distribution. An example trajectory starting at $x_0=(0,0)$ and ending at $x_2$, running in the time interval $[0,2]$ is depicted using a white to red colour gradient indicating the progression of time. The periodic nature of the diffusion can be seen by the wrapping of both the stationary diffusion and the trajectory at the boundaries of the square plane. The fact that stationary distribution is not aligned with the horizontal and vertical axes illustrates the dependence (given by $\alpha_3$) between the $\phi$ and $\psi$ dihedral angles.}
\label{fig:wnd}
\end{figure}

Topologically, the WN diffusion (see Figure~\ref{fig:wnd} for a pictorial example) can be thought of as the analogue of the OU process on the torus $\mathbb{T}^2=[-\pi,\pi)\times[-\pi,\pi)$. The WN diffusion arises as the wrapping on $\mathbb{T}^2$ of the following Euclidean diffusion:
\begin{align}
dX_{t}=\overbrace{A\underset{k\in\mathbb{Z}^{2}}{\sum}(\mu-X_{t}-2k\pi)w_{k}(X_{t})}^{\substack{\text{drift}\\\text{coefficient}}}dt\; +\overbrace{\Sigma^{\frac{1}{2}}}^{\substack{\text{diffusion}\\\text{coefficient}}}dW_{t},
\label{eq:wnd}
\end{align}
where $W_{t}$ is the two-dimensional Wiener process, $A$ is the drift matrix, $\mu\in\mathbb{T}^{2}$ is the stationary mean, $\Sigma$ is the infinitesimal covariance matrix and
\begin{equation}
w_{k}(\theta)=\frac{\phi_{\frac{1}{2}A^{-1}\Sigma}(\theta-\mu+2k\pi)}{\underset{m\in\mathbb{Z}^{2}}{\sum}\phi_{\frac{1}{2}A^{-1}\Sigma}(\theta-\mu+2m\pi)},
\label{eq:weighting}
\end{equation}
$\theta\in\mathbb{T}^{2}$, is a probability density function (pdf) for $k\in\mathbb{Z}^{2}$. $\phi_\Sigma$ stands for the pdf of a bivariate Gaussian $\mathcal{N}(0,\Sigma)$. The pdf \eqref{eq:weighting} weights the linear drifts of \eqref{eq:wnd} such that they become smooth and periodic.\\

It is shown in \citet{garciap2017diffusions} that the stationary distribution of the WN diffusion is a $\mathrm{WN}(\mu,\Sigma)$, which has pdf:
\begin{equation}
p_{\mathrm{WN}}(\theta|\mu,\Sigma)=\sum_{k\in\mathbb{Z}^{2}}\phi_{\Sigma}(\theta-\mu+2k\pi).\label{eq:wnstatdist}
\end{equation}
Despite involving an infinite sum over $\mathbb{Z}^{2}$, taking just the first few terms of this sum provides a tractable and accurate approximation to the stationary density for most of the realistic parameter values.\\

Maximum Likelihood Estimation (MLE) for diffusions is based on the transition probability density (tpd), which only has a tractable analytical form for very few specific processes. A highly tractable  and accurate approximation to the tpd is given for the WN diffusion. This approximation results from weighting the tpd of the OU process in the same fashion as the linear drifts are weighted in \eqref{eq:wnd}, yielding the following multimodal pseudo-tpd:
\begin{align}
\tilde{p}(\theta_2|\theta_1,A,\mu,\Sigma,t)=\sum_{m\in\mathbb{Z}^{2}}p_{\mathrm{WN}}(\theta_2|\mu_{t}^{m},\Gamma_{t})w_{m}(\theta_2),\label{eq:ap}
\end{align}
with $\theta_1,\theta_2\in\mathbb{T}^2$, $\mu^m_{t}=\mu+e^{-tA}(\theta_1-\mu+2\pi m)$ and $\Gamma_{t}=\int_{0}^{t}e^{-sA}\Sigma e^{-sA^{T}}ds$. The pseudo-tpd provides a good approximation to the true tpd in key circumstances: \textit{i}) $t\to0$, since it collapses in the Dirac's delta; \textit{ii}) $t\to\infty$, since it converges to the stationary distribution; \textit{iii}) high concentration, since the WN diffusion becomes an OU process. Furthermore, it is shown in \citet{garciap2017diffusions} that the pseudo-tpd has a lower Kullback--Leibler divergence with respect to the true tpd than the Euler and Shoji--Ozaki pseudo-tpds, for most typical scenarios and discretization times in\nopagebreak[4] the diffusion trajectory.\\

A further desirable property of the pseudo-tpd is that it obeys the time-reversibility equation, which in terms of $(X_a^{x(i)}, X_b^{y(i)})$ is
\begin{align*}
\tilde{p}(X_b^{y(i)}|X_a^{x(i)},A,\mu,\Sigma,t_{ab})p_{\mathrm{WN}}(X_a^{x(i)}|\mu,\tfrac{1}{2}A^{-1}\Sigma)=\tilde{p}(X_a^{x(i)}|X_b^{y(i)};A,\mu,\Sigma,t_{ab})p_{\mathrm{WN}}(X_b^{y(i)}|\mu,\tfrac{1}{2}A^{-1}\Sigma).
\end{align*}
Indeed, the WN diffusion is the \textit{unique} time-reversible diffusion with the stationary pdf \eqref{eq:wnstatdist}, in the same way the OU is with respect to a Gaussian. Time-reversibility is an assumption of the overall model and many other models of sequence evolution. A benefit of time-reversibility in a pairwise model such as ETDBN is that one of the proteins in a pair may be arbitrarily chosen as the ancestor, thus avoiding a computationally expensive marginalisation of an unobserved ancestor.\\

The likelihood of a dihedral angle observation pair $(X_a^{x(i)},X_b^{y(i)})$, assuming that $X_a^{x(i)}$ is drawn from the stationary distribution, is given by:
\begin{align}
p(X_{a}^{x(i)},X_{b}^{y(i)}|H^{i},r_{a}^{i},r_{b}^{i},t_{ab})&=p(X_a^{x(i)},X_b^{y(i)}|A,\mu,\Sigma,t_{ab})\notag\\
&\approx\tilde{p}(X_b^{y(i)}|X_a^{x(i)},A,\mu,\Sigma,t_{ab})  p_{\mathrm{WN}}(X_a^{x(i)}|\mu,\tfrac{1}{2}A^{-1}\Sigma),
\label{eq:wnlikelihood}
\end{align}
$A$ and $\Sigma$ are constrained to yield a covariance matrix $A^{-1}\Sigma$. A parametrization that achieves this is $\Sigma=\mathrm{diag}(\sigma_{1}^{2},\sigma_{2}^{2})$ and $A=\big(\alpha_{1}, \tfrac{\sigma_{1}}{\sigma_{2}}\alpha_{3}; \tfrac{\sigma_{2}}{\sigma_{1}}\alpha_{3}, \alpha_{2}\big)$, $\alpha_{1}\alpha_{2}>\alpha_{3}^{2}$. $\alpha_{1}$ and $\alpha_{2}$ are the drift components for the $\phi$ and $\psi$ dihedral angles, respectively. Dependence (correlation) between the dihedral angles is captured by $\alpha_{3}$. A depiction of a WN diffusion with given drift and diffusion parameters is depicted \nolinebreak[4] in Figure~\ref{fig:wnd}.\\

The computationally tractable nature of the stationary density and tpd required in \eqref{eq:wnlikelihood} is key to enabling efficient training of and sampling under the model. Specifically, the computation of \eqref{eq:ap} involves evaluating $e^{-tA}$ and $\Gamma_{t}$, which we can work out explicitly. First, $e^{-tA}=a(t)I-b(t)A$ with $a(t)=e^{-rt}(\cosh(qt)+r\frac{\sinh(qt)}{q})$, $b(t)=e^{-rt}\frac{\sinh(qt)}{q}$, $r=\tfrac{\mathrm{tr}(A)}{2}$ and $q=\sqrt{|\det(A-rI)|}$. Second, since $A^{-1}\Sigma$ is symmetric,
\[
\Gamma_{t}=s(t)\tfrac{1}{2}A^{-1}\Sigma+i(t)\Sigma,
\]
with $s(t)=1-a(2t)$ and $i(t)=\tfrac{b(2t)}{2}$. This gives a neat interpolation of the stationary and infinitesimal covariance matrices, particularly convenient for efficiently evaluating \eqref{eq:ap} at different\nolinebreak[4] $t$'s.

\subsection{Evolutionary regimes: modelling shift and drift}

We now turn to the meaning of the evolutionary regime pairs. Two modes of evolution are modelled: \textit{constant evolution} and \textit{jump evolution}. Constant evolution occurs when the evolutionary regime starting in protein $p_a$ at aligned site $i$, $r_{a}^{i}$, is the same as the evolutionary regime ending in protein $p_b$  at aligned site $i$, $r_{b}^{i}$, \textit{i.e.} $r_{a}^{i}=r_{b}^{i}$. Conversely, jump evolution occurs when $r_{a}^{i}\neq r_{b}^{i}$. Constant evolution is intended to capture angular drift (changes in dihedral angles localised to a region of the Ramachandran plot), whereas jump evolution is intended to capture angular shift (large changes in dihedral angles, possibly between distant regions of the Ramachandran plot).\\

The hidden state at node $H^{i}$, together with the evolutionary time $t_{ab}$ separating proteins $p_{a}$ and $p_{b}$, specifies a joint distribution over the evolutionary regime pairs:
\begin{equation}
p(r_{a}^{i},r_{b}^{i}|H^{i},t_{ab})=p(r_{a}^{i}|H^i, r_{b}^{i},t_{ab})p(r_{b}^i|H^i),
\label{eq:jump_jointlikelihood}
\end{equation}
where
\begin{align*}
p(r_{a}^{i}|H^{i}, r_{b}^{i},t_{ab})=\begin{cases}
e^{-\gamma_{H^{i}}t_{ab}}+\pi_{H^{i},r_{b}^{i}}(1-e^{-\gamma_{H^{i}}t_{ab}}), & \text{if }r_{a}^{i}=r_{b}^{i},\\
\pi_{H^{i},r_{b}^{i}}(1-e^{-\gamma_{H^{i}}t_{ab}}), & \text{if }r_{a}^{i}\neq r_{b}^{i},\\
\end{cases}
\end{align*}
and
$p(r_{a}^{i}|H^{i})=\pi_{H^{i},r_{a}^{i}}$ and $p(r_{b}^{i}|H^{i})=\pi_{H^{i},r_{b}^{i}}$. $\pi_{H^{i},r_{a}^{i}}$ and $\pi_{H^{i},r_{b}^{i}}$ are model parameters specifying the probability of starting in regime $r_{a}^{i}$ or $r_{b}^{i}$, respectively, corresponding to the hidden state specified by node $H^{i}$. $\gamma_{H^{i}} > 0$  is a model parameter specifying the jump rate corresponding to the hidden state specified by node $H^{i}$.\\

The regime pair jump probabilities have been chosen so that time-reversibility holds, in other words:
\begin{align*}
p(r_{a}^{i}|H^i, r_{b}^{i},t_{ab})p(r_{b}^{i}|H^i)=p(r_{b}^{i}|H^i,r_{a}^{i},t_{ab})p(r_{a}^{i}|H^i).
\end{align*}
The hidden state at node $H^i$, together with a regime pair $(r_{a}^{i},r_{b}^{i})$ and the evolutionary time $t_{ab}$, specifies the joint likelihood over site observation pairs:
\begin{align}
p&(O_{a}^{x(i)},O_{b}^{y(i)}|H^{i},r_{a}^{i},r_{b}^{i},t_{ab})=\begin{cases}
p(O_{a}^{x(i)},O_{b}^{y(i)}|H^{i},r_{c}^{i},t_{ab}), & \text{if }r_{a}^{i}=r_{b}^{i}=r_{c}^i,\\
p(O_{a}^{x(i)}|H^{i},r_{a}^{i})p(O_{b}^{y(i)}|H^{i},r_{b}^{i}), & \text{if }r_{a}^{i}\neq r_{b}^{i}.\\
\end{cases}
\label{eq:likelihood}
\end{align}
In the case of constant evolution, evolution at aligned $i$  is described in terms of the same evolutionary regime $r_{c}^{i}$. Evolution is considered constant because each observation type is drawn from a single stochastic process specified by $H^i$ and $r_c$. Note that the strength of the evolutionary dependency within an observation pair is a function of the evolutionary time $t_{ab}$.\\

In the case of jump evolution, the evolutionary processes are, after the evolutionary jump, restarted independently in the stationary distribution of the new evolutionary regime. Thus the site observations $O_{a}^{x(i)}$ and $O_{b}^{y(i)}$ are assumed to be drawn from the stationary distributions of two separate stochastic processes corresponding to evolutionary regimes $r_{a}^{i}$ and $r_{b}^{i}$, respectively. This implies that, conditional on a jump, the likelihood of the observations is no longer dependent on $t_{ab}$. Furthermore, there is no longer an evolutionary trajectory linking the two site observations, hence there is no need to perform a computationally expensive marginalisation over all possible trajectories, as might be necessary in a model with continuous-time Markovian switching between evolution regimes.
The likelihood of an observation pair is now simply a sum over the four possible regime pairs:
\begin{align*}
p(O^{x(i)},O^{y(i)}|H^{i},t_{ab})=\sum_{(r_{a}^{i},r_{b}^{i})\in{R}} p(O^{x(i)},O^{y(i)}|H^{i},r_{a}^{i},r_{b}^{i},t_{ab}) p(r_{a}^{i},r_{b}^{i}|H^{i},t_{ab}),
\end{align*}
where $R=\{(1,1),(1,2),(2,1),(2,2)\}$ is the set of four regime pairs, $p(O^{x(i)},O^{y(i)}|H^{i},r_{a}^{i},r_{b}^{i})$ is given by (\ref{eq:likelihood}) and $p(r_{a}^{i},r_{b}^{i}|H^{i},t_{ab})$ is given by (\ref{eq:jump_jointlikelihood}).

\subsection{Identification of evolutionary motifs encoding jump evolution}
\label{sec:idingevomotifs}

In order to identify aligned sites having potential evolutionary motifs encoding jump evolution, a specific criterion was developed.\\

For a particular protein pair, inference was performed under the model conditioned on the amino acid sequence and dihedral angles for both proteins, $(A_a,A_b,X_a,X_b)$. Homologous sites corresponding to a single hidden state and with evidence of jump evolution ($r_a^{i}\neq r_b^{i}$) at posterior probability $>0.90$ were identified, that is, the $i$'s such that $p(H^i,r_a^{i}\neq r_b^{i}|A_a,A_b,X_a,X_b)>0.90$.\\

In a second filtering step, amino acid sequences and a single set of dihedral angles corresponding to one of the proteins were used ($A_a,A_b,X_a$ or $A_a,A_b,X_b$) to infer the posterior probability, this time at a lower threshold: $p(H^i,r_a^{i}\neq r_b^{i}|A_a,A_b,X_a)>0.50$ or $p(H^i,r_a^{i}\neq r_b^{i}|A_a,A_b,X_b)>0.50$. This second criterion ensured that the evolutionary motif was identifiable under typical conditions where one has limited access to structural information (in this case a single protein structure in a pair).

Only those aligned sites meeting both criteria were selected for further downstream analysis.

\subsection{Statistical alignment: modelling insertions and deletions}

Protein sequences can not only undergo amino transitions due to underlying nucleotide mutations in the coding sequence, but also indel events. To account for indels, a modified pairwise TKF92 alignment HMM  based on \citet{miklos2004long} was implemented. The TKF92 alignment HMM was augmented with additional evolutionary hidden states intended to capture local sequence and structure evolutionary dependencies. Furthermore, it was modified such that neighbouring dependencies amongst hidden states at adjacent alignment sites were modelled.\\

Whilst it is possible to fix the alignment in advance by pre-aligning the sequences using one of the many available alignment methods (Katoh et al. (2002); Edgar (2004)) or using a curated alignment (such as from the HOMSTRAD database), doing so ignores alignment uncertainty. For more detail we refer to the supplementary material.

\subsection{Training and test dataset}

A training dataset of 1200 protein pairs (2400 proteins; 417,870 site observation pairs) and a test dataset of 38 protein pairs (76 proteins; 14125 site observation pairs) were assembled from 1032 protein families in the HOMSTRAD database.\\

For each protein family in HOMSTRAD (ranging in size from 2 to 22 homologous proteins each), a phylogenetic tree was inferred from the HOMSTRAD protein family sequence alignment using FastTree \citep{price2010fasttree}. Each protein family tree was taken and protein pairs selected such that the sum of the branches between pairs  was maximised, whilst ensuring that no pair of proteins in the set shared an overlapping evolutionary history. This was done in order to maximise the amount of information, whilst minimising dependencies due to shared evolutionary history.\\

Dihedral angles were computed from the PDB coordinates of each protein structure using the BioPython.PDB package \citep{hamelryck2003pdb}. Furthermore, each protein was taken and the corresponding full length protein structure obtained from the PDB database and the secondary structure annotated at each amino acid position using DSSP \citep{touw2015series}. For some proteins we were not able to obtain the full length PDB structures, which may result in misclassification of secondary structure interactions due to truncation. For these structures the secondary structure annotations were treated as missing. All 38 proteins in the test dataset had secondary structure annotations and were from distinct protein families.

\subsection{Model training}

Maximum likelihood estimation of the model parameters, $\hat{\Psi}$, was done using Stochastic Expectation Maximisation (StEM).
Forward Filtering Backward Sampling (FFBS) was used to jointly sample alignment configurations ($M_{ab}$), hidden node states ($H$) and evolutionary regimes, ($r_{a},r_{b}$). The Metropolis--Hastings algorithm was used to sample the four pair-specific continuous parameters $\theta_{ab}=\{t_{ab},\lambda_{ab},\mu_{ab},r_{ab}\}$: evolutionary time ($t_{ab}$), insertion ($\lambda_{ab}$), deletion ($\mu_{ab}$) and geometric-extension ($r_{ab}$) rates. In other words, at iteration $k$ for each pair of unaligned observation sequences $O_{a}$ and $O_{b}$ we draw samples, from the following joint-distribution:
\begin{align*}
Z_{ab}^{(k)}\sim p(M_{ab},H,r_{a},r_{b},\theta_{ab}|O_{a},O_{b},\Psi^{(k)}).
\end{align*}

In the M-step the samples from the previous E-step, were used to update the hidden node parameters ($\hat{\Psi}$) using efficient sufficient statistics and COBYLA optimization algorithm \citep{powell1994cobyla} in the NLOpt library \citep{johnson2014nlopt} was used.

\subsection{Angular distances}

For benchmarking purposes, the angular cosine distance was used to measure distances between pairs of dihedral angles, $\langle\phi_{a},\psi_{a}\rangle$ and $\langle\phi_{b},\psi_{b}\rangle$. It is defined as follows \citep{downs2002circular}:
\begin{align}
d&(\langle\phi_{a},\psi_{a}\rangle,\langle\phi_{b},\psi_{b}\rangle)=\sqrt{4-2\cos(\phi_{a}-\phi_{b})-2\cos(\psi_{a}-\psi_{b})}.
\label{eq:angular_distance}
\end{align}
The maximum possible distance is $\sqrt{8} \approx 2.828$. It has the property that when $\phi_{a}-\phi_{b}\approx 0$ and $\psi_{a}-\psi_{b}\approx 0$ are near zero it may be approximated by the Euclidean distance -- using the small angle approximation for cosine ($\cos\theta\approx1-\frac{\theta^{2}}{2}$  when $\theta$ is near zero):
\begin{align*}
d(\langle\phi_{a},\psi_{a}\rangle,\langle\phi_{b},\psi_{b}\rangle) \approx\sqrt{4-2(1-(\phi_{a}-\phi_{b})^{2}/2)-2(1-(\psi_{a}-\psi_{b})^{2}/2)} = \sqrt{(\phi_{a}-\phi_{b})^{2}+(\psi_{a}-\psi_{b})^{2}}.
\end{align*}

\section{Results and discussion}

\subsection{Selecting the number of hidden states}

Models with 8, 16, 32, 48, 64, 80, 96 and 112 hidden states were trained until convergence for three different repetitions using different initial random number seeds. The highest log-likelihood model of the three repetitions for each number of hidden states was selected for downstream analysis.\\

Following that, marginal likelihoods $p(D|\text{model})$, \textit{i.e.} model evidence  for each of the 38 protein pairs in the test dataset, were computed under each model by fixing the alignments to the respective HOMSTRAD alignments. The alignments were fixed \textit{a priori} in order to make computation of the marginal likelihoods computationally tractable. Additionally, predictive accuracies under a homology modelling scenario, $p(X_b|A_a,A_b,X_a,\text{model})$, were calculated for the same 38 protein pairs.\\

The 112 hidden state model had the highest total marginal log-likelihood for protein pairs in the test dataset and predictive accuracy under the homology modelling scenario comparable to the predictive accuracies of the 16, 32, 48, 64, 80 and 96 hidden state models (all were with standard error of the mean, see supplementary data). Only the model with 8 hidden states had significantly poorer predictive accuracy on the test dataset.\\

\begin{figure}[h!]
\iffigures
\centering
\includegraphics[width=0.9\textwidth]{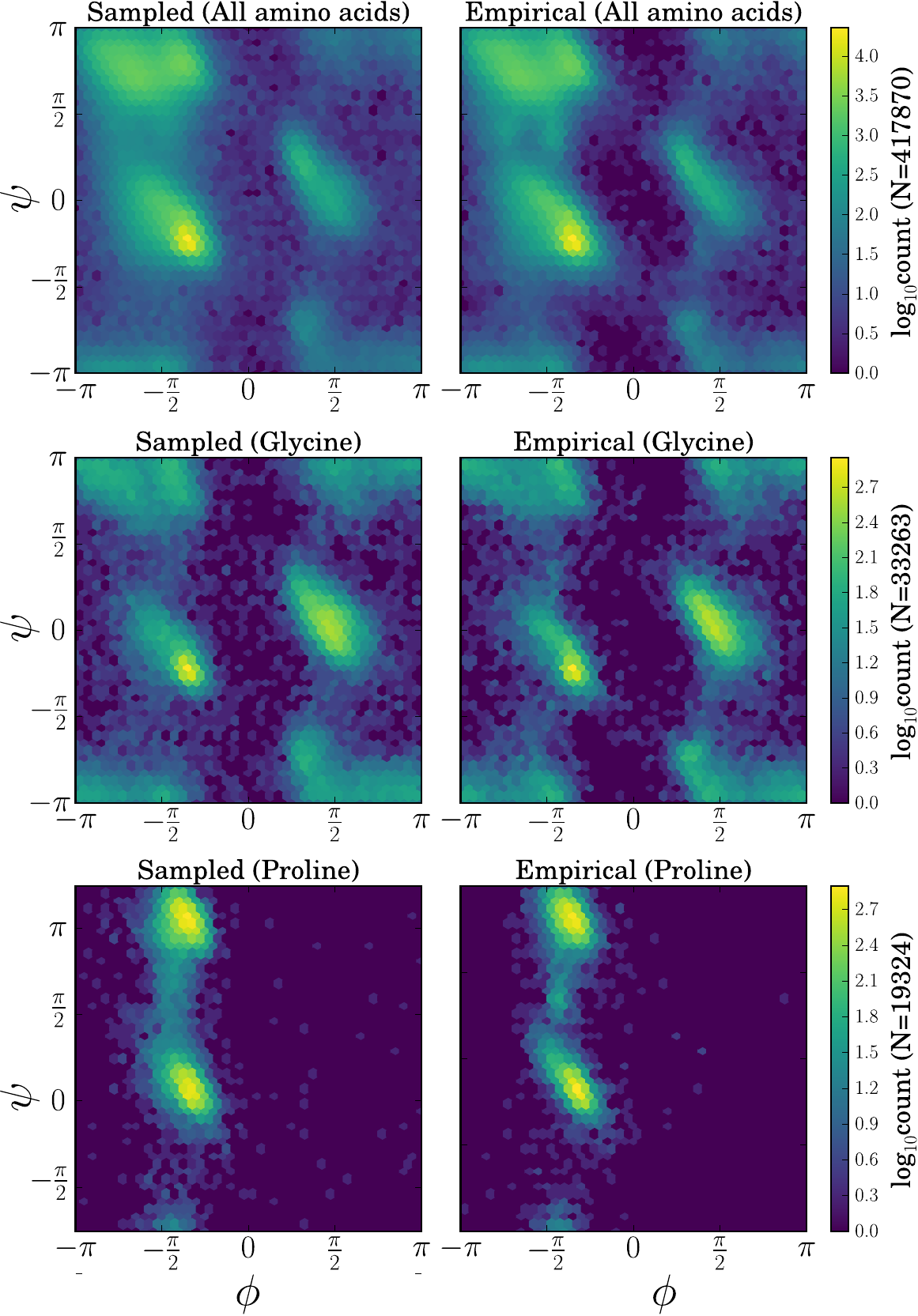}
\caption{\small Ramachandran plots depicting  sampled and empirical dihedral angle distributions. The top row depicts the distributions for all amino acids, the middle for glycine only and the bottom for proline only. The leftmost plots show dihedral angles sampled under the jump model, whereas the rightmost plots show the empirical distributions of dihedral angles in the training dataset.}
\fi
\label{fig:stationary_distributions}
\end{figure}

Although the model with 112 hidden states had the highest total marginal log-likelihoods, the model with 64 hidden states was selected as the final model. A 64 hidden state model was chosen as a trade-off between total marginal log-likelihood on the test dataset, computational time and the number of model parameters. Inference under the model scales $\mathcal{O}(h^2)$ in the number of hidden states and the higher total marginal log-likelihood of the 112 hidden state model did not justify the increase in computational time when sampling from the model.

\subsection{Stationary distributions over dihedral angles capture the empirical distribution}

Figure~\ref{fig:stationary_distributions} illustrates the sampled and empirical dihedral angle distributions. There is a good correspondence between dihedral angles sampled under the model (Figure~\ref{fig:stationary_distributions}, left) and the empirical distribution of dihedral angles in our training dataset (Figure~\ref{fig:stationary_distributions}, right) for all three cases illustrated (all amino acids, glycine only and proline only).
The correspondence is not surprising given that ETDBN is effectively a mixture model with a large number of mixture components, however, it is a good indication that there are a lack of obvious errors in the model implementation.

\subsection{Estimates of evolutionary time from dihedral angles are consistent with estimates from sequence}

Figure~\ref{fig:branch_length_scatter} compares evolutionary times estimated using only pairs of homologous amino acid sequences only versus pairs of homologous dihedral angles only. As desired, the two estimates of evolutionary time for each protein pair are similar, as can be seen by the proximity of the points to the identity\nolinebreak[4] line.\\

A paired $t$-test gave a $p$-value of 0.578, thus failing to reject the null hypothesis that there is no difference between branch lengths estimated using sequence only vs. angles only. This indicates that there is sufficient evolutionary information in the dihedral angles to estimate the evolutionary times and that the model is consistent in its estimates, lacking a significant tendency to under-estimate or over-estimate the evolutionary times when either sequence or dihedral angles are used.\\

Interestingly, the variance in the sampled evolutionary times is higher when dihedral angles only are used, as compared to sequence only (see Figure \ref{fig:alignmenthmm}).

\begin{figure}[t!]
	\iffigures
	\centering
	\includegraphics[width=0.55\textwidth]{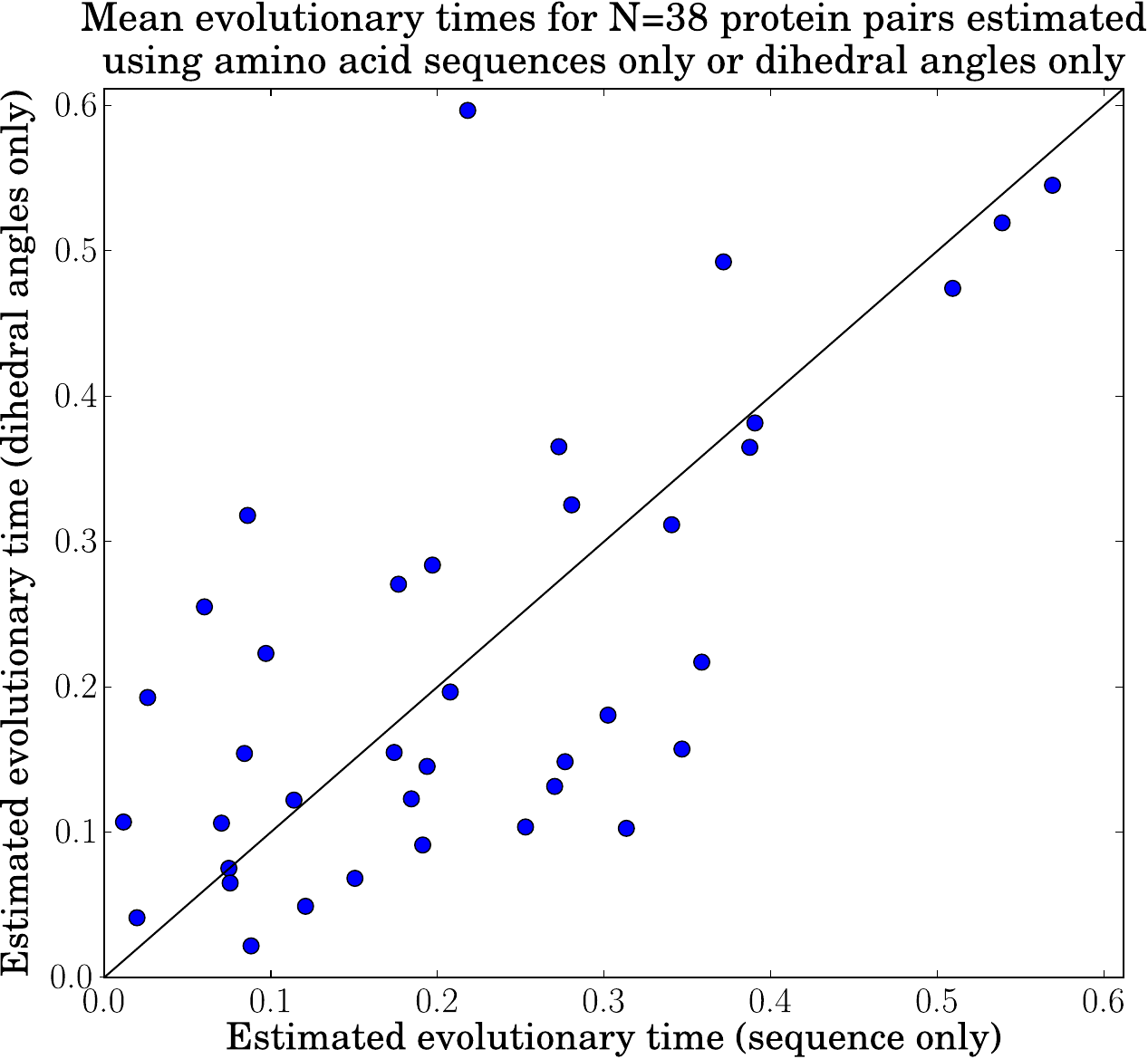}
	\fi
	\caption{\small Scatterplot comparing evolutionary times estimated using pairs of homologous amino acid sequences only versus pairs of homologous sets of dihedral angles only for $N=38$ proteins pairs in the test dataset. The $x$-coordinate of each point gives the estimated evolutionary time based only on the amino acid sequence, whereas the $y$-coordinate gives the estimated evolutionary time based only on the dihedral angles. The diagonal line represents $y=x$.}
	\label{fig:branch_length_scatter}%
\end{figure}

\subsection{The relationship between evolutionary time and angular distance is adequately modelled}

We investigated the relationship between evolutionary time and angular distance between real protein pairs and protein pairs where the dihedral angles of $p_b$ ($X_b$) were treated as missing and hence sampled (Figure~\ref{fig:evolutionarytimeangulardistance}).\\

As expected, for both real and sampled pairs, angular distance tends to increase as a function evolutionary time. For larger evolutionary times a plateau begins to emerge, which is expected as the maximum possible theoretical angular distance is $\sqrt{8}\approx2.828$.\\

When the evolutionary time is exactly zero ($t_{ab}=0$) under our model, the angular distance between sampled dihedral angles is exactly zero (not shown in Figure~\ref{fig:evolutionarytimeangulardistance}), however, this is not expected to be the case for real protein pairs when the two sequences are identical (due to the inherently flexible nature of proteins, different experimental conditions, experimental noise, etc.). It is therefore not surprising that the regression curve for the real protein pairs does not pass through zero.\\

For small evolutionary times ($<0.2$) the curves for the real and sampled protein pairs show a good correspondence, however, for larger evolutionary times the model tends to under-estimate angular distances. This likely reflects the fact that the tpd of the WN diffusion specified is localised around its mean when $t_{ab} \to \infty$ and therefore dihedral angles distant from this mean are unlikely to be sampled. To a certain extent this is mitigated by the jump model, which occasionally allows for large changes in dihedral angle, but is still somewhat limited in its flexibility, as it only allows jumps between two evolutionary regimes.

\begin{figure}[h!]
	\iffigures
	\centering
	\includegraphics[width=0.55\textwidth]{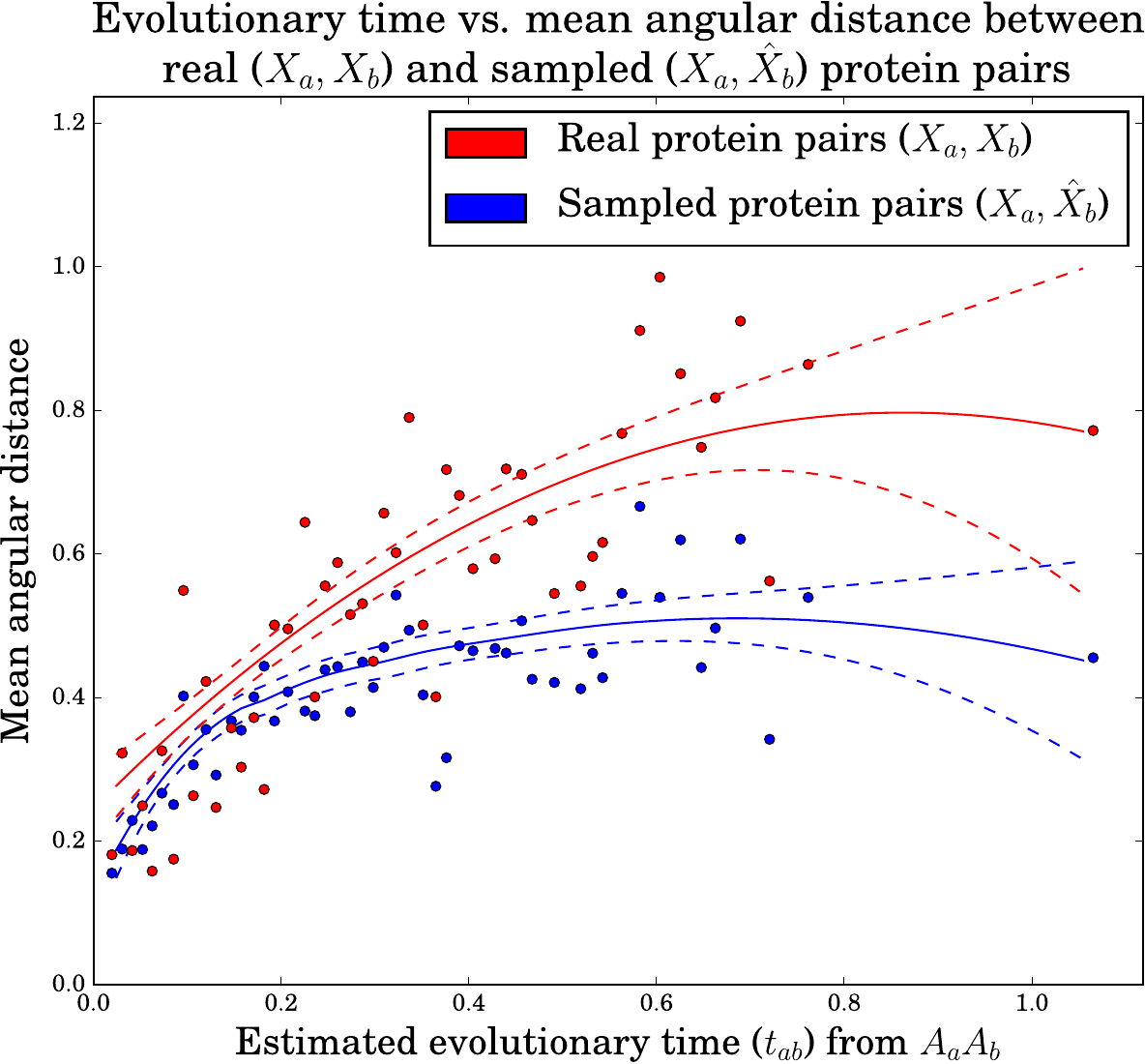}
	\fi
	\caption{\small Evolutionary time vs. angular distances between real and corresponding sampled proteins pairs in the training dataset at 50 representative evolutionary times. Mean angular distances between the real dihedral angles in a protein pair (red, $X_a$ and $X_b$) and sampled dihedral angles in a sampled protein pair (blue, $X_a$ and $\hat{X}_b$) were compared to test how well the sampled dihedral angles reproduced the real angular distances. The dihedral angles ($\hat{X}_b$) of each sampled protein pair were sampled by conditioning on both amino acid sequences and the homologous dihedral angles ($A_a$, $A_b$, $X_a$), and the estimated evolutionary time ($\hat{t}_{ab}$) for the real protein pair. The regression curves were obtained by a quadratic LOcally-weighted regrESSion (LOESS), with smoothing parameter chosen by leave-one-out cross-validation. The 95\% confidence intervals for the mean assume error normality.}
	\label{fig:evolutionarytimeangulardistance}%
\end{figure}

\subsection{Evaluation of the model}

The conditional independence structure in \eqref{eq:conditional_independence} enables computationally efficient sampling from the model under different combinations of observed or missing data. For example, ETDBN can be used to sample (\textit{i.e.} predict) the dihedral angles of a protein from its corresponding amino acid sequence, a homologous amino acid sequence, a homologous set of dihedral angles, the corresponding secondary structure, a homologous secondary structure or any combination of them.\\

Predictive accuracy was measured using 38 homologous protein pairs in the test dataset. For every protein pair $(p_a,p_b)$, the dihedral angles of $p_b$ in each pair were treated as missing, and these missing dihedral angles were sampled under the model given a particular combination of observation types. The average angular distance (\ref{eq:angular_distance}) between the sampled and known dihedral angles was used as the measure of predictive accuracy.\\

Figure~\ref{fig:benchmarks_single} gives an example of predictive accuracy under different combinations of observations types overlaid on a cartoon structure of the protein structure being predicted, whereas Figure~\ref{fig:benchmarksa} provides a representative view of predictive accuracy across 10 different protein pairs in the test dataset for different combinations of observations types. We highlight some the key patterns identified in Figures~\ref{fig:benchmarks_single} and~\ref{fig:benchmarksa} here. \nowidow

\begin{figure}[h!]
	\iffigures
	\centering
	\includegraphics[width=0.9\textwidth]{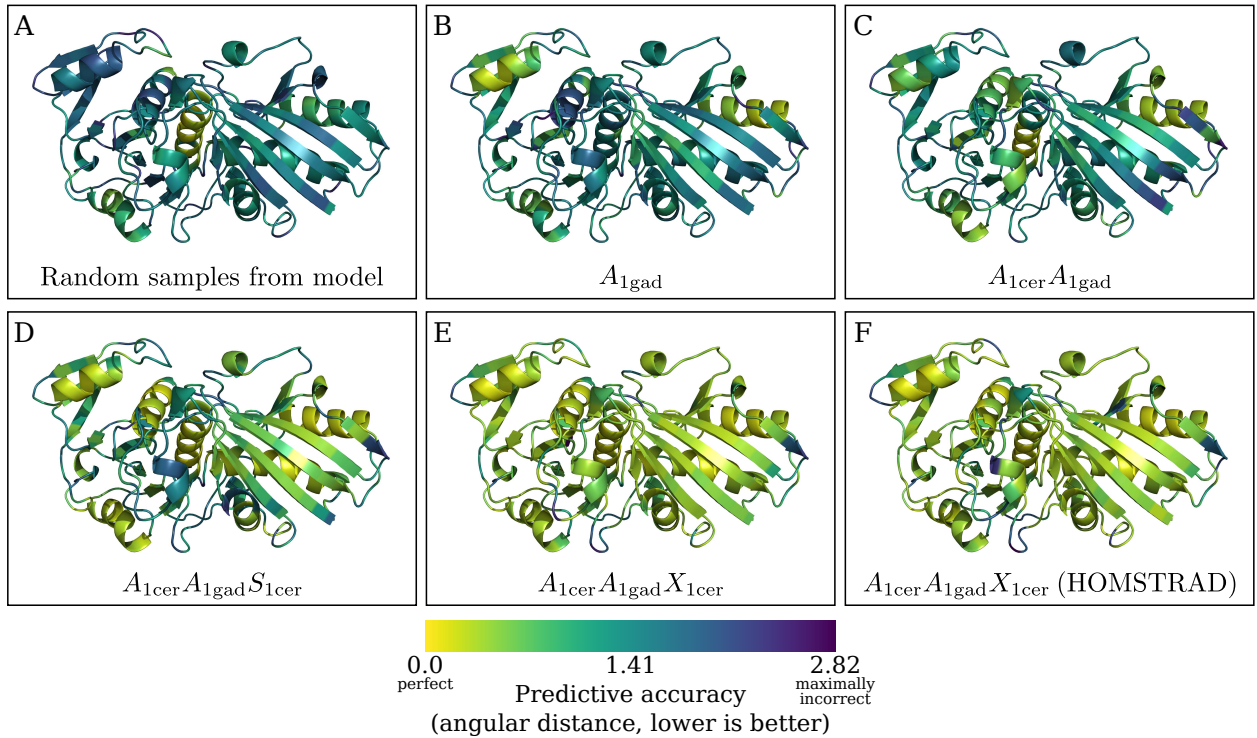}
	\fi
	\caption{\small Cartoon structure representations of \textit{E.coli} glyceraldehyde-3-phosphate dehydrogenase structure (PDB 1gad) are depicted in each panel, overlaid with predictive accuracy when using different combinations of observed data  to predict missing dihedral angles in 1gad. \textit{Thermus aquaticus} glyceraldehyde-3-phosphate dehydrogenase (PDB 1cer) was used as a homolog for the purposes of prediction. Predictive accuracy is indicated using a colour gradient depicting the mean angular distance between the true dihedral angle ($X_{\text{1gad}}^{i}$) and the predicted (sampled) dihedral angles ($\hat{X}_{\text{1gad}}^{i}$) at each amino acid position. The label at the bottom of each panel indicates the data combination used. In A, no data was used for prediction. In B, only the amino acid sequence corresponding to 1gad ($A_{\text{1gad}}$) was used. In C, the amino acid sequence of 1gad ($A_{\text{1gad}}$) and the amino acid sequence of the homologous protein ($A_{\text{1cer}}$) were used. In D, both amino acids sequences ($A_{\text{1cer}}$ and $A_{\text{1gad}}$) and the secondary structure of the homologous protein ($S_{\text{1cer}}$) were used. In E, both the amino acid sequences ($A_{\text{1cer}}$ and $A_{\text{1gad}}$) and the dihedral angles of the homologous protein ($X_{\text{1cer}}$)  were used. Finally, in panel F the same combination of observations was used as in E, but the alignment was treated as known \textit{a priori}.}
	\label{fig:benchmarks_single}
\end{figure}

Combination 1 refers to random sampling from the model, implying no data observations were conditioned on besides the respective lengths of proteins $p_a$ and $p_b$. The average angular distance between the true and predicted dihedral angles was 1.6. Random sampling acts as a baseline for predictive accuracy. It is apparent from Figure~\ref{fig:benchmarks_single} that the model has a propensity to predict right-handed $\alpha$-helices, which is the most populated region in the Ramachandran plot.\\

Under combination 2, only the amino acid sequence corresponding to $p_b$ is observed. As expected in Figures~\ref{fig:benchmarks_single} and~\ref{fig:benchmarksa} there is an increase in predictive accuracy with the addition of the amino acid sequence relative to combination 1.\\

Under combination 3, we add in the amino acid sequence of a homologous protein ($p_a$). In all ten cases there is an improvement in predictive accuracy. The improvement in predictive accuracy is reasonable, as knowledge of the sequence evolutionary trajectory is expected to encode information about structure evolution and hence will inform the dihedral angle conformational possibilities.\\

Under combination 4, in addition to the two amino acid sequences we treat the homologous secondary structure as observed. This results in a substantial improvement in predictive accuracy as one would expect. Knowledge of the amino acid sequence and a homologous secondary structure strongly informs regions of the Ramachandran plot that are likely to be occupied. \noclub[3] 

\begin{figure}[h!]
	\iffigures
	\centering
	\includegraphics[width=0.9\textwidth]{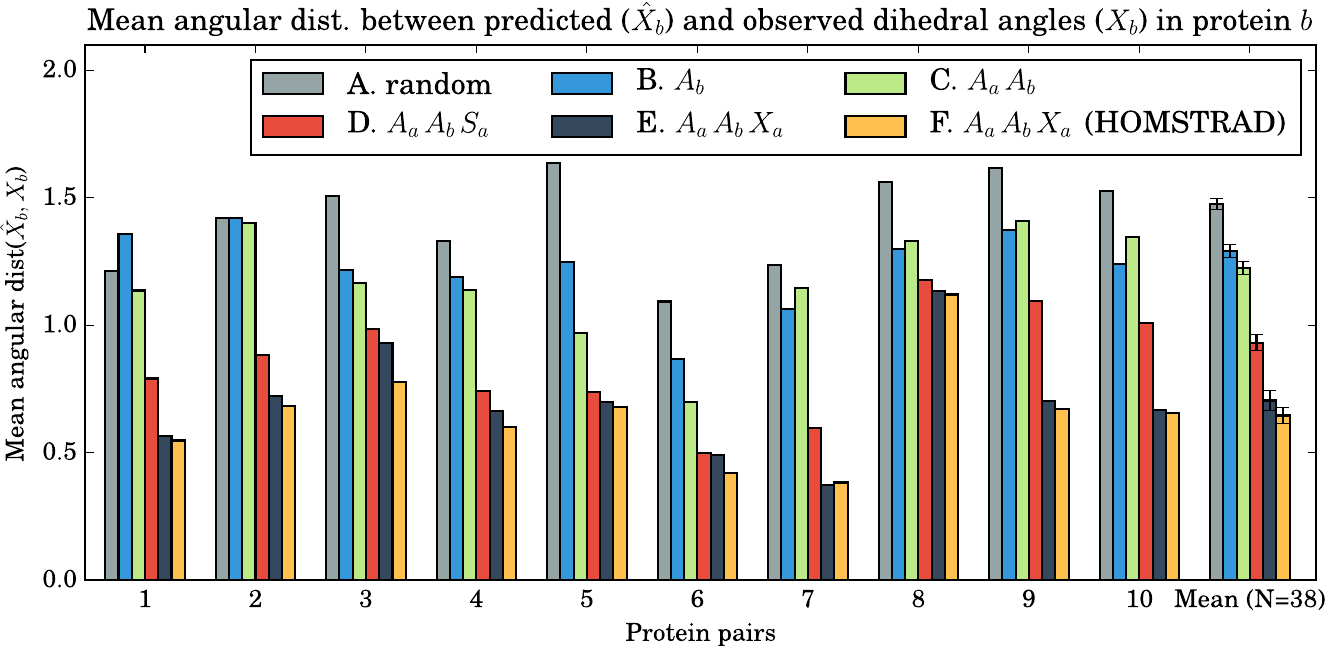}
	\fi
	\caption{\small Benchmarks of predictive accuracy (measured using angular distance, lower is better) on a random subset of ten protein pairs in the test dataset, giving a representative view of predictive accuracy under six different combinations of observations. The dihedral angles $X_b$ of protein $b$ were treated as missing and were sampled under the model, whereas $p_a$ was a homologous protein used for the purposes of prediction. See the legend of Figure~\ref{fig:benchmarks_single} for a description of each combination (A-F). The final set of bars, denoted `Mean (N=38)', are the mean values for the entire test dataset of N=38 protein pairs. The error bars are the standard errors.}
	\label{fig:benchmarksa}%
\end{figure}

Under combination 5 (which we consider the \textit{canonical} combination -- the standard homology modelling scenario), we treat both amino acid sequences as observed, as well as the dihedral angles of the homologous protein ($p_a$) -- in all cases the predictive accuracy improves over combination 4. This is anticipated as the homologous dihedral angles are expected to be the best proxy for missing dihedral angles and are therefore expected to be more informative than secondary structure alone. Note that the availability of a homologous amino acid sequence pair here and in combination 4 is consequential as it informs the evolutionary time $t_{ab}$ parameter, which will typically constrain the distribution over dihedral angles and reduce the associated uncertainty.\\

Finally, in combination 6, the same data observations as in combination 5 are used, except the alignment is treated as given \textit{a priori} (by the HOMSTRAD alignment) rather than as unobserved. The HOMSTRAD alignment is based on a structural and sequence alignment of $p_a$ and $p_b$ and therefore is expected to encode a higher degree of homology and structural information than combination 5 (where the alignment is treated as unobserved and therefore a marginalisation over alignments is performed). On average, there is a slight improvement in predictive accuracy when fixing the alignment, albeit, the magnitude of improvement is not substantial. This demonstrates the correctness of the alignment HMM.\\

The alignment HMM is valuable accounts for alignment uncertainty in a principled manner, this is particularly useful when an appropriate alignment is unavailable. However, it should be noted that inference scales $\mathcal{O}(|p_{a}||p_{b}|h^2)$ when treating the alignment as unobserved, where $|p_{a}|$ and $|p_{b}|$ are the lengths of $p_a$  and $p_b$, respectively. Inference scales $\mathcal{O}(mh^2)$ when the alignment is fixed \textit{a priori}, where $m$ is the length of alignment $M_{ab}$ and is typically much smaller than $|p_{a}||p_{b}|$.\\

It should be emphasised that we do not expect ETDBN to compete with structure prediction packages such as Rosetta \citep{rohl2004protein} or homology modelling software such as \citet{arnold2006swiss} in terms of predictive accuracy. Our current model is a local model of structure evolution -- it is not even expected capture constraints such as the radius of gyration of a protein or other global features typical of proteins.

\subsection{Evolutionary hidden states reveal a common evolutionary motif}

One benefit of ETDBN  is that the 64 evolutionary hidden states learned during the training phase are interpretable. We give an example of a hidden state encoding jump evolution that was subsequently found to represent an \textit{evolutionary motif} present in a large number of protein pairs in our test and training datasets.\\

Evolutionary hidden state 3 (Figure~\ref{fig:hiddenstate3}) was selected from the 64 hidden states as an example of a hidden state encoding jump evolution and capturing angular shift (a large change in dihedral angle). A notable feature of this hidden state is that the change in dihedral angles between evolutionary regimes $r_1$ and $r_2$ is associated with specific amino acid changes. In regime $r_1$ the amino acid frequencies are relatively spread out amongst a number of amino acids, whereas in regime $r_2$ the frequencies are particularly concentrated in favour of glycine (Gly) and asparagine (Asp), with glycine being significantly more probable in regime $r_2$ than $r_1$. This suggests that conditioned on hidden state 3, an exchange between a glycine to another amino acid is likely indicative of a jump and hence a corresponding change in dihedral angle. This is consistent with what we find in a subsequent analysis of evolutionary motifs. This particular jump occurs in coil regions.\\

Having selected hidden state 3, positions in 238 protein pairs were analysed for evidence of the corresponding evolutionary motif. 38 protein pairs in the test dataset and a further 200 from the training dataset were analysed using the criteria described in the Methods section. Using the first criterion, 84 protein sites in 59 protein pairs corresponding to $H^i=3$ (evolutionary hidden state 3) were identified. Of the 84 protein sites, 34 protein sites met the second criterion.\\

We give an example of a homologous protein pair illustrating the identified evolutionary motif. Two histidine-containing phosphocarriers, 1pch (\textit{M. capricolum}) and 1poh (\textit{E. coli}), were identified as having the evolutionary motif (Figure~\ref{fig:hpr}) at homologous site E39/G39.\\

Most positions in the homologous pair have very low jump probabilities ($\approx0.0$), with the exception of positions N38/N38 and E39/G39, which both have high posterior jump probabilities ($\approx1.0$). The exchange between a glutamate (at position 39 in 1poh) and a glycine (at position 39 in 1pch) appears to be responsible for the shift in dihedral angle. This exchange corresponds to a significant jump in dihedral angle ($\langle\phi_{\text{1poh,E39}}=-1.63,\psi_{\text{1pch,E39}}=-0.06\rangle\rightarrow\langle\phi_{\text{1poh,G39}}=1.40,\psi_{\text{1pch,G39}}=0.22\rangle$). The angular distance between the two dihedral angles is 2.01. This is consistent with the amino acid frequency parameters specified by the two regimes for hidden state 3 (Figure~\ref{fig:hiddenstate3}).\\

\begin{figure}[h!]
	\iffigures
	\centering
	\includegraphics[width=0.9\textwidth]{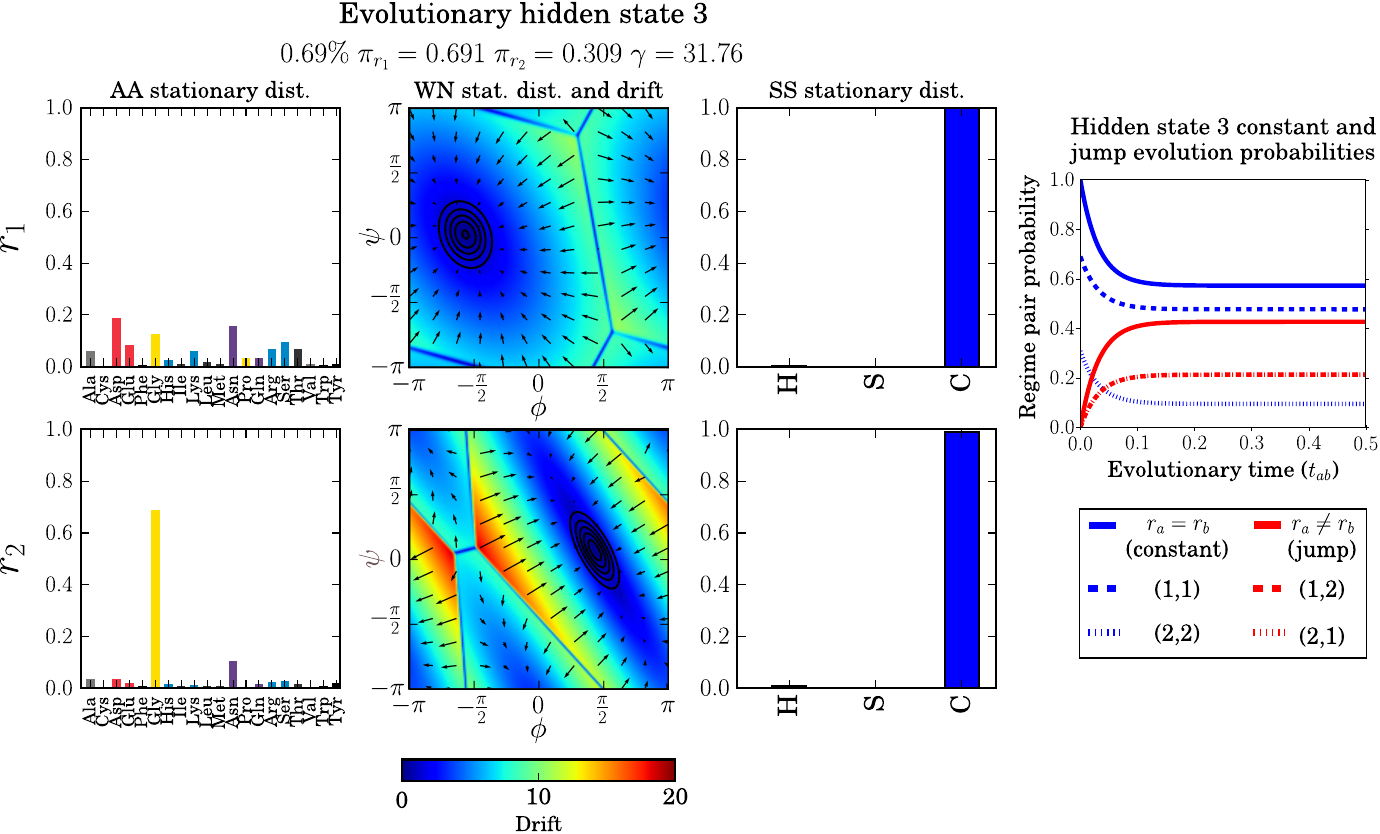}
	\fi
	\caption{\small Depiction of evolutionary hidden state 3. This hidden state was sampled at $0.69\%$ of sites (the average was $1.56\%$). The equilibrium frequencies of $r_1$ and $r_2$ were ${\pi_1}=0.691$ and ${\pi_2}=0.309$, respectively. The jump rate was $\gamma=31.76$. The corresponding regime pair probabilities are depicted to the right as a function of evolutionary time. In the main figure, the two rows depict the parameters encoded by the two evolutionary regimes, respectively. Columns 1  and 3 depict the parameters governing the the amino acid and secondary structure stochastic processes, respectively. The secondary structure classes correspond to H=helix, S=sheet and C=coil. Column 2 depicts the WN diffusions. The stationary distributions of the WN diffusions are shown using black contour lines, the direction of the drifts are indicated by the arrows and the magnitude of the drifts at each position indicated using the colour gradient.}
	\label{fig:hiddenstate3}%
\end{figure}

Regime $r_1$ indicates that a number of amino acids (alanine, aspartic acid, glycine, histidine, lysine, asparagine, proline, glutamine, arginine, serine and theorine) other than glutamate plausibly coincide with the particular dihedral angle conformation specified by regime $r_1$. The involvement of glycine in a jump is not surprising as it is a small and flexible amino acid, whereas the role of asparagine is less clear. In our analysis of 238 protein pairs we found that of the seven positions meeting the criteria for hidden state 3 and involving an exchange with asparagine (Asn), four were an exchange between an asparagine and a glycine, whereas the remaining three were between asparagine and one of lysine, histidine or serine.

\subsection{Using dihedral angles for alignment}

A valuable feature of our model is its ability to account for alignment uncertainty by summing over possible pairwise alignments using the TKF92 model as a prior distribution over indel histories, whilst simultaneously taking into account neighbouring dependencies amongst aligned sites. Doing so results in a sample of alignments rather than a single alignment. Nevertheless, a single Maximum A Posteriori (MAP) pairwise alignment may be obtained from the alignment samples and used for downstream analysis.\\

ETDBN and several other alignment methods (namely StatAlign, BAli-Phy, MUSCLE and MAFFT) were used to infer pairwise alignments from simulated and real data under various combinations of data observations, for example: an amino acid sequence pair ($A_a,A_b$), a secondary structure sequence pair ($S_a,S_b$), a dihedral angle sequence pair ($X_a,X_b$) and combinations thereof.\\

In the first set of benchmarks (Figure~\ref{fig:alignmentsim}A), pairs of proteins were simulated from the ETDBN model conditioned on 38 different pairwise alignments and corresponding evolutionary times. This resulted in a set of 38 simulated pairwise alignments together with corresponding observations, implying that the true underlying alignments were known for each of the simulated protein pairs. ETDBN and a number other alignment methods were used to infer pairwise alignments for each. The alignment similarity metric \citep{schwartz2005alignment} was used to measure the similarity between the inferred alignments and the true alignments, where higher similarity indicates better predictions. It was found that, when using the simulated amino acid sequences alone, ETDBN (\ref{fig:alignmentsim}A.5) outperformed all four other methods tested (\ref{fig:alignmentsim}A.1 MUSCLE, \ref{fig:alignmentsim}A.2 MAFFT, \ref{fig:alignmentsim}A.3 StatAlign, \ref{fig:alignmentsim}A.4 BAli-Phy). The greater performance of ETDBN compared to other methods can not be considered a fair comparison, as the data were simulated under the ETDBN model.\\

\begin{figure}[h!]
	\iffigures
	\centering
	\includegraphics[width=0.8\textwidth]{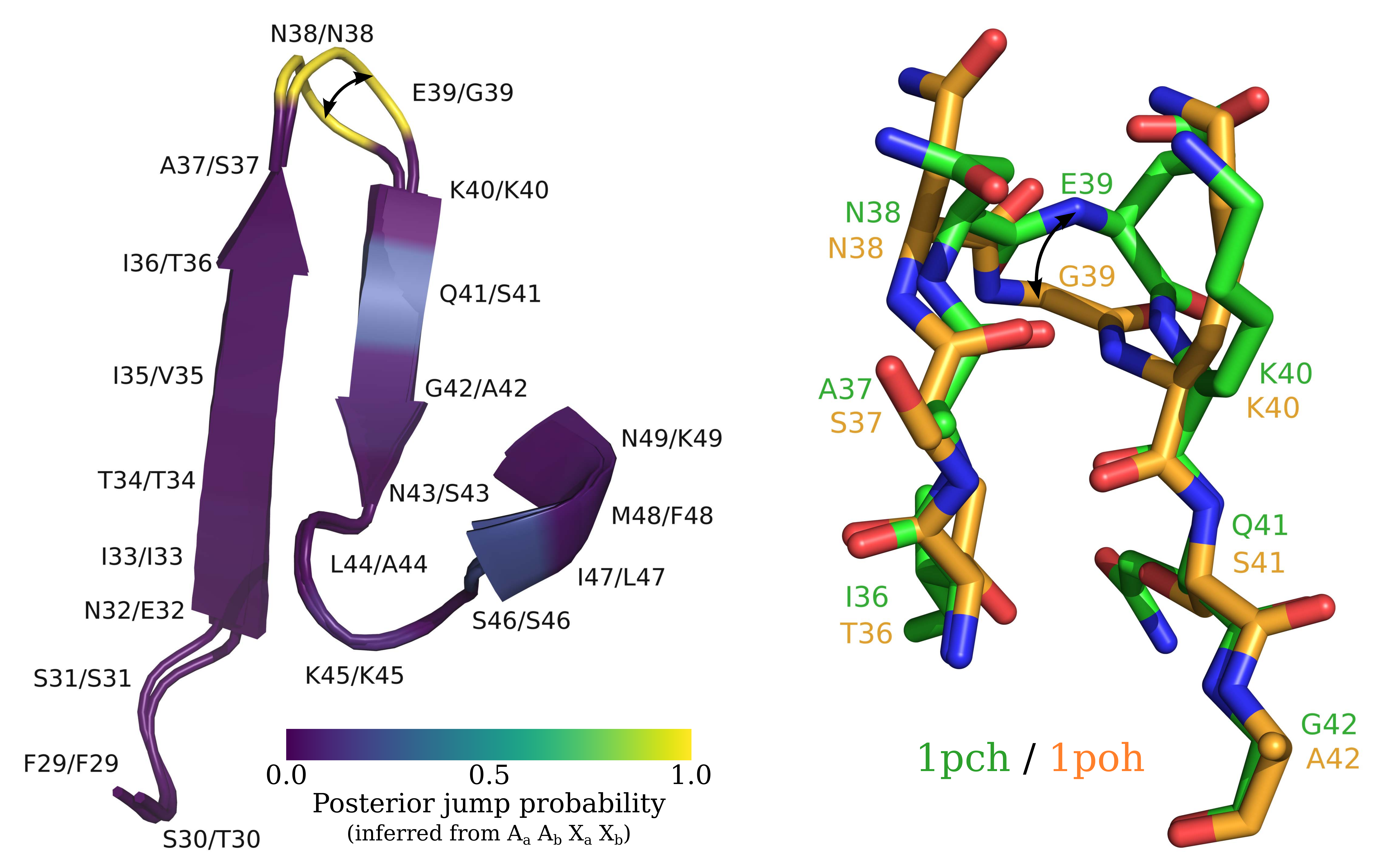}
	\fi
	\caption{\small Depiction of two histidine-containing phosphocarriers, PDB 1pch and 1poh, superimposed. On the left is a cartoon representation of the two proteins corresponding to regions F29-K49 and F29-A42, respectively, with posterior jump probabilities at each position overlaid. On the right is a ball-and-stick representation giving atomic detail for a smaller region (I36-G42 and T36-A42, respectively). The exchange between a glutamate (E39 in 1poh) and a glycine (G39 in 1pch) is associated with a large change in dihedral angle as indicated by the curved arrows.}
	\label{fig:hpr}%
\end{figure}

\begin{figure}
	\iffigures
	\centering
	\includegraphics[width=0.45\textwidth]{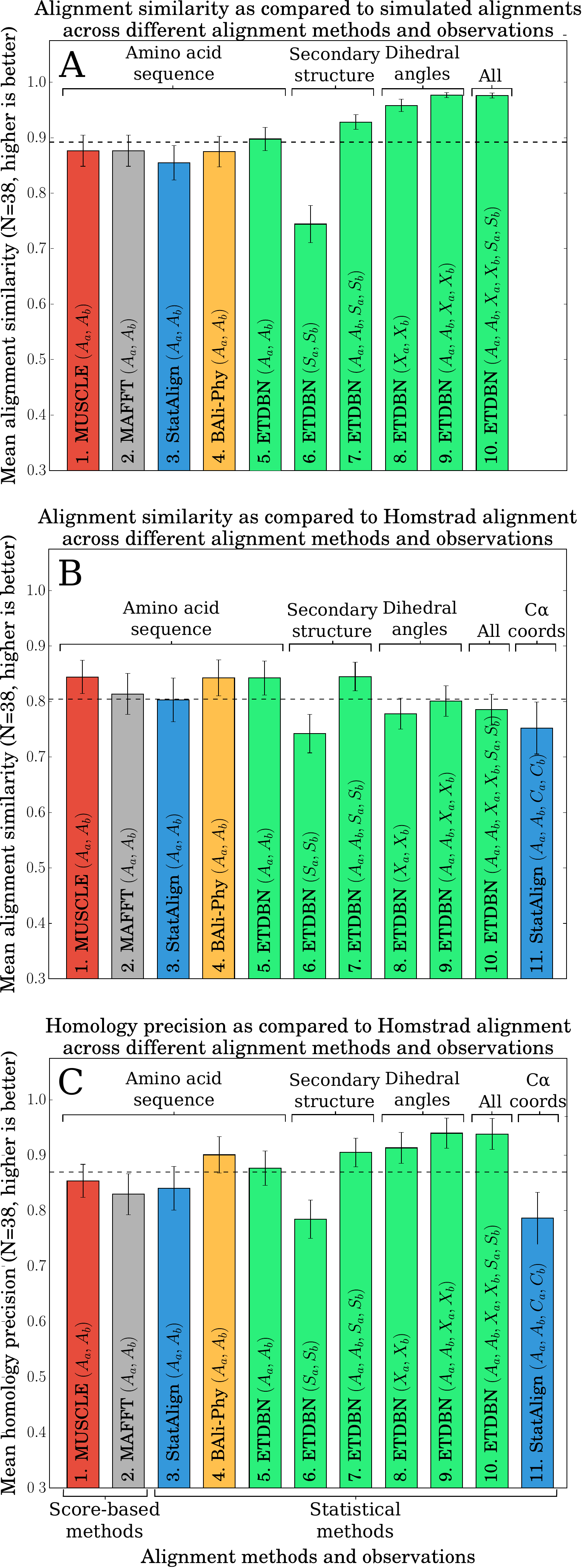}
	\fi
	\caption{\small Alignment benchmarks. A: averages of alignment similarity across different methods and combinations of data observations, where the observations were simulated from the ETDBN model conditioned on a single alignment, hence the underlying alignment was known. B: averages of alignment similarity, where the 38 HOMSTRAD alignments in the test dataset were taken as the true alignments. C: averages of precision in predicting homologous site pairs in 38 sequence pairs from the test dataset, where homologous site pairs in the HOMSTRAD alignments were taken to be the true homologous site pairs. Note: combination 11 in panels B and C denotes that the StatAlign structural alignment method \citep{herman2014simultaneous} was used to align the two proteins based on the $C\alpha$ coordinates.}
	\label{fig:alignmentsim}%
\end{figure}

More revealing in Figure~\ref{fig:alignmentsim}A was the alignment similarity under ETDBN when using different combinations of simulated data observations. It was found that secondary structure alone (\ref{fig:alignmentsim}A.6) performed the worst, which is unsurprising given that only three states were available to align the proteins. The second worst in terms of alignment similarity was amino acid sequences alone (\ref{fig:alignmentsim}A.5), followed by amino acid sequences  and secondary structures (\ref{fig:alignmentsim}A.7). Interestingly, using dihedral angles only (\ref{fig:alignmentsim}A.8) outperformed both \ref{fig:alignmentsim}A.5 (sequences only) and \ref{fig:alignmentsim}A.6 (secondary structures only). Finally, using amino acid sequence together with dihedral angles (\ref{fig:alignmentsim}A.9) or all three data types combined (\ref{fig:alignmentsim}A.10) outperformed all other combinations. This illustrates that, at least under simulation conditions, increasing the number of data observations results in better alignment \linebreak accuracy.\\

Following that, the various alignment methods were benchmarked against 38 pairwise alignments consisting of real sequence and structure observations in the test dataset. These pairwise alignments were obtained from the HOMSTRAD alignments. The sequence identity of these pairwise alignments ranged from $10\%$ to $93\%$, with an average sequence identity of $39\%$. When benchmarking the MAP estimated alignments against the HOMSTRAD alignments (Figure~\ref{fig:alignmentsim}B), using real sequences alone for inference $(A_a,A_b)$, ETDBN (\ref{fig:alignmentsim}B.5) had a similar degree of accuracy when compared to several other sequence-based methods (\ref{fig:alignmentsim}B.1 StatAlign, \ref{fig:alignmentsim}B.2 BaliPhy, \ref{fig:alignmentsim}B.3 MUSCLE and \ref{fig:alignmentsim}B.4 MAFFT). This demonstrates that ETDBN has performance comparable to that of other commonly-used sequence alignment methods.\\

Using $(S_a,S_b)$  alone, ETDBN (\ref{fig:alignmentsim}B.6) had substantially lower alignment similarity compared to sequence only, which was expected given that a similar result was obtained for the simulated data (\ref{fig:alignmentsim}A.6). However, when including the real sequences (\ref{fig:alignmentsim}B.7) the predictive accuracy was once again comparable to sequence only inferences (\ref{fig:alignmentsim}B.1--\ref{fig:alignmentsim}B.5).\\\noclub[3] 

When using ($X_a X_b$) alone (\ref{fig:alignmentsim}B.8), the alignment similarity was found to be somewhat worse than the sequence only cases. Furthermore, when introducing the sequences (\ref{fig:alignmentsim}.9) and secondary structures (\ref{fig:alignmentsim}B.10) in addition to the dihedral angles, the similarity still remained worse than the sequence only methods (\ref{fig:alignmentsim}B.1--\ref{fig:alignmentsim}B.5), despite the additional information. These results are in contrast to the results we obtained for simulated data (\ref{fig:alignmentsim}A.8--\ref{fig:alignmentsim}A.10).\\

Upon further investigation, the trend was found to reverse (Figure~\ref{fig:alignmentsim}C) when the precision of predicting homologous sites was calculated (the fraction of sites which were predicted as homologous and were correctly predicted as such). Therefore when only dihedral angle observations are used, ETDBN underpredicts the number of homologous sites, however, when a homologous site is predicted, it is correctly predicted more often than when using only amino acid sequences. In particular, ETDBN predicted fewer homologous sites with coiled secondary structure compared to homologous sites with helical or sheet secondary structure. This pattern of results may be in part due to the WN diffusion used to model evolution of dihedral angles. The WN diffusion is suitable for modelling angular drift (small changes in angle localised around a region of the Ramachandran plot) but does not sufficiently capture angular shift (large changes in angle between regions of the Ramachandran plot, which are more likely in coiled regions). The jump model to some extent captures angular shift, but it can only do so for a single evolutionary jump between two regions of the Ramachandran plot. In other words, it does not consider multiple different homologous angular shift events caused by amino acid mutation.\\

When interpreting these results it important to note the HOMSTRAD alignments should not be considered the true underlying alignments and may even be strongly biased (e.g. they may favour the most parsimonious alignments, with the fewest number of indels). In practice, it is extremely difficult to obtain the true underlying alignment, because it would require an experiment where every indel event since the common ancestor is observed, a seeming impossible task outside of simulation or laboratory conditions.

\section{Concluding remarks}
\label{sec:Conclusion}

The main achievement of this work is a computationally tractable, generative and interpretable probabilistic model of protein sequence and structure evolution on a local scale.\\

Previous stochastic models of protein sequence and structure evolution emphasised estimation of evolutionary parameters \citep{challis2012stochastic,herman2014simultaneous}. ETDBN is somewhat of a departure from these previous models, but is likewise capable of estimating evolutionary parameters. We show that estimates of evolutionary times inferred under ETDBN are consistent regardless of whether amino acid sequence or dihedral angle observations are used. In addition, the relationship between evolutionary time and angular distance in real proteins is adequately recapitulated in protein pairs sampled under the model, albeit the angular distance is under-estimated for larger evolutionary times, which might be explained by the limited flexibility of the jump model. Like previous models, ETDBN is capable of dealing with alignment uncertainty by marginalising over indel histories; it predicts pairwise MAP consensus alignments with accuracy similar to that of score-based and statistical alignment methods.\\

The generative nature of ETDBN allows us to demonstrate that the underlying empirical distributions over dihedral angles (depicted using Ramachandran plots) are captured and that the model is capable of predicting missing observations, such as dihedral angles, from a variety of different data types. For example, an amino acid sequence, a homologous amino acid sequence, a homologous secondary structure, a homologous set of dihedral angles or any combination thereof.\\

Due to its local nature, ETDBN does not constitute a homology modelling method in itself. Rather, it can be used as a building block, much like fragment libraries model local structure in protein structure prediction methods. ETDBN places the homology modelling problem on a statistical footing, enabling a number of approaches to later be used, such as multi-level modelling, \textit{i.e.} combining fine-grained distributions (for example, distributions over dihedral angles, such as ETDBN) and coarse-grained distributions (for example, distributions describing the global properties of proteins, such as compactness).\\

In addition to multi-level modelling, probabilistic models such as ETDBN allow one to account for and to make statements about uncertainty (e.g. with respect to evolutionary time, alignment, etc.) in a rigorous manner. One immediate use of ETDBN from a structure prediction or homology modelling perspective is as an efficient proposal distribution. ETDBN could be used to sample protein structures (possibly conditioned on various data observations) in a highly computationally efficient manner, such that the resulting samples are expected to be located in regions of high probability density with respect to the true underlying distribution.\\

A final key feature of our evolutionary model is its interpretable nature. This interpretability enables the identification of potential evolutionary motifs -- common patterns of sequence-structure evolution. We identify one such evolutionary motif in 34 different homologous protein pairs. A major direction for future research is the further identification of such evolutionary motifs. Understanding these evolutionary motifs, may \textit{i}) improve homology modelling predictions; \textit{ii}) provide more accurate estimates of evolutionary parameters; and \textit{iii}) produce better models of protein evolution that more realistically capture evolutionary trajectories through sequence and structure space, which may help identify functionally relevant positions that may be possible drug targets.

\section{Future challenges}

For reasons of computational tractability the implemented model is pairwise, but it is theoretically possible to generalise it to a phylogeny, such as in \citet{herman2014simultaneous}. In practice, for three or more sequences on a phylogeny it is necessary to marginalise out the unobserved ancestral protein states in order to compute likelihoods. Felsenstein's algorithm can be used to marginalise over discrete ancestral states, such as amino acids in a computationally efficient manner. However, we have not yet established whether a similar efficient algorithm exists for marginalising the continuous ancestral dihedral angle states under the WN diffusion, thereby necessitating a more expensive MCMC algorithm. A possibly greater computational hindrance to considering a phylogeny is the alignment problem, which scales $\mathcal{O}(l_{1}\times{l_{2}}\times\cdots\times{l_{N}})$, where $l_i$ is the length of sequence $i$ and $N$ is the number of sequences, although MCMC approaches are also possible \citep{herman2014simultaneous}.\\

Although we believe our model provides a substantial improvement over current stochastic models of sequence and structural evolution, there is still scope for improvement. The WN diffusions used to model dihedral angle evolution adequately capture angular drift (small local changes in dihedral angle), but are less capable of capturing angular shift (large changes in dihedral angle). This problem is somewhat mitigated by our jump model, but even this only allows jumps between two possible evolutionary regimes conditioned on a particular hidden state. A more realistic model would allow an arbitrary number of switches between evolutionary regimes together with neighbouring dependencies amongst adjacent sites along the evolutionary trajectory. Continuous-Time Bayesian Networks \citep{nodelman2002continuous} provide such a framework for large state space models with certain sparsity conditions, but will likely incur a large penalty in terms of computational efficiency.

\section*{Software availability}

Julia code (tested on both Windows and Linux platforms) is available upon request from the first author. 

\section*{Supplement}

A supplement contains further details on the methods employed in the paper and additional results.

\section*{Acknowledgments}

The authors acknowledge funding from the University of Copenhagen 2016 Excellence Programme for Interdisciplinary Research (UCPH2016-DSIN). The second author acknowledges support from project MTM2016-76969-P from the Spanish Ministry of Economy and Competitiveness and ERDF. Authors acknowledge valuable comments from three referees that led to substantial improvements in the manuscript, as well as initial discussions with Christian Ravn, Ian Lim and Mathias Cronj\"ager.


\fi

\ifsupplement

\appendix
\newpage
\title{Supplement to ``A generative angular model of protein structure evolution''}
\setlength{\droptitle}{-1cm}
\predate{}%
\postdate{}%
\date{}

\author{Michael Golden$^{1,6}$, Eduardo Garc\'ia-Portugu\'es$^{2,3}$, Michael S\o rensen$^{3}$,\\ Kanti V.~Mardia$^{4}$, Thomas Hamelryck$^{2,5}$, and Jotun Hein$^{1}$}

\footnotetext[1]{
Department of Statistics, University of Oxford (UK).}
\footnotetext[2]{
Bioinformatics Centre, Section for Computational and RNA Biology, Department of Biology, University of Copenhagen (Denmark).}
\footnotetext[3]{
Department of Mathematical Sciences, University of Copenhagen (Denmark).}
\footnotetext[4]{
Department of Mathematics, University of Leeds (UK).}
\footnotetext[5]{
Image Section, Department of Computer Science, University of Copenhagen (Denmark).}
\footnotetext[6]{Corresponding author. e-mail: \href{mailto:golden@stats.ox.ac.uk}{golden@stats.ox.ac.uk}.}
\maketitle

\begin{abstract}
This supplement contains further details on the methods employed in the paper and additional results.
\end{abstract}
\begin{flushleft}
	\small\textbf{Keywords:} Evolution; Protein structure; Probabilistic model; Directional statistics.
\end{flushleft}

\section{Supplementary methods}

\subsection{Statistical alignment: modelling insertions and deletions with neighbouring dependencies}

Protein sequences can not only undergo point mutation events, but also also insertion and deletion (indel) events. We describe a modified pairwise TKF92 alignment HMM that models both local sequence/structure evolution and sequence alignment based on \citep{miklos2004long}.\\

Whilst it is possible to fix the alignment in advance by pre-aligning the sequences using one of the many available optimisation-based alignment methods \citep{katoh2002mafft,edgar2004muscle} or using a curated alignment (such as from the HOMSTRAD database), doing so ignores alignment uncertainty.\\

An alignment can be thought of as a statement about homology, such that when amino acid positions are aligned in order to indicate homology they are considered to have evolved solely via mutation along the evolutionary trajectory linking them and therefore not arising via an indel. As the evolutionary trajectory of indels is rarely observed in practice, it is difficult to make statements about the true underlying alignment (homology relationships), especially when the sequences being compared are distantly related and/or the rate indel of evolution is high. The TKF92 model \citep{thorne1992inching} gives a suitable distribution describing indel evolution.\\

For the pairwise case, the TKF92 model can be represented as an HMM using the formulation described in Mikl\'{o}s et al. (2008). This HMM formulation allows one to sum over all possible pairwise alignments in $\mathcal{O}(nm)$ time using the HMM forward-backward algorithm, where $n$ and $m$ are the respective lengths of the two sequences. Thereby accounting for alignment uncertainty due to insertions or deletions.\\

\begin{figure}[h]
	\centering
	\iffigures
	\includegraphics[width=0.9\textwidth]{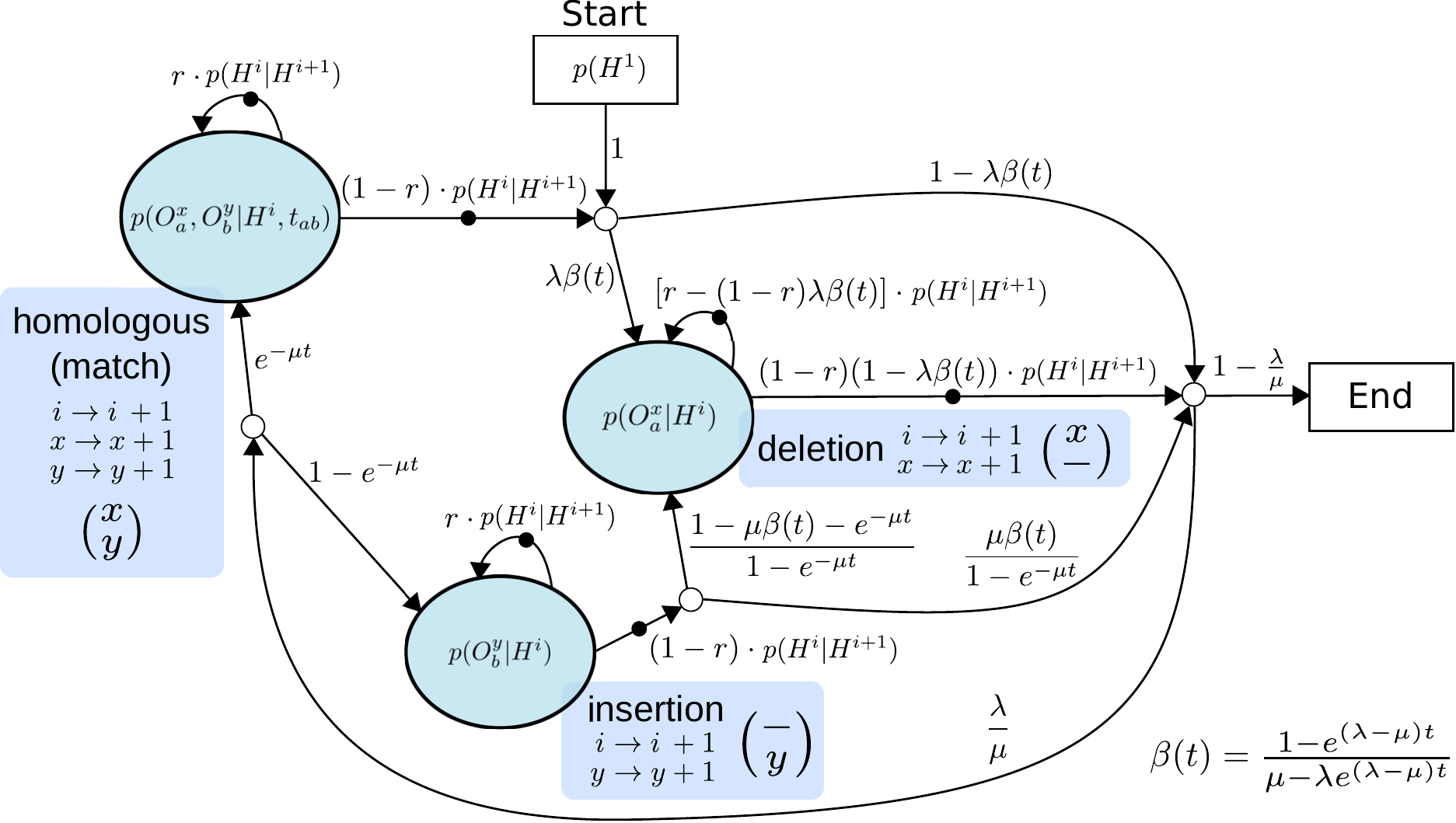}
	\fi
	\caption{\small Diagrammatic representation of modified TKF92 alignment with neighbour-dependent evolutionary hidden states. The small white nodes represent non-emitting nodes. The three large oval nodes represent emitting nodes, emitting a deletion with respect to the first sequence $\binom{x}{-}$, a insertion with respect to the first sequence $\binom{-}{y}$ or a homologous pair of amino acids $\binom{x}{y}$. Edges with filled black circles indicate that the evolutionary hidden state is permitted to transition to a potentially different evolutionary hidden state, such transitions are permissible on all edges which exit one of the three emitting states. Parameters determining the transition probabilities are as follows: the insertion rate ($\lambda>0$),  the deletion rate ($\mu>0$, the extension rate ($0<r<1$) and the evolutionary time $t_{ab}$. It is required that $\lambda < \mu$, such that the distribution ($t\rightarrow\infty$) of sequence lengths is finite at stationarity. The evolutionary hidden state transition probabilities, $P(H^i|H^{i-1})$, are given by a $h \times h$ transition probability matrix and the initial probabilities at the first alignment site, $p(H^i)$, is given by a length $h$ probability vector.}
	\label{fig:alignmenthmm}%
\end{figure}

We implemented a modified version of this HMM implementation (Figure~\ref{fig:alignmenthmm}) such that each emitted pair of characters is drawn from one of $h$ evolutionary hidden states (Figure~\ref{fig:alignmenthmm}). Additionally, we encode neighbouring dependencies amongst evolutionary hidden states along the alignment, by specifying a probability transition matrix $p(H^{i},H^{i+1})$, that allows the hidden states to transition at ``Insertion'', ``Deletion'' or ``Match'' nodes. These hidden states are intended to encode local sequence and structure evolution. The introduction of evolutionary hidden states with neighbouring dependencies increases the computational complexity from $\mathcal{O}(nm)$ in a model without evolutionary hidden states to $\mathcal{O}(nmh^{2})$ in a model with evolutionary hidden states, where $h$ is the number of evolutionary hidden states. The likelihood of observation pair under this model depends on the homology relationship, $M_{ab}^{i}$, at a given position is as follows:
\begin{align*}
p(O^{i}|M_{ab}^{i},H^{i},t_{ab})=\begin{cases}
p(O_{a}^{x(i)},O_{b}^{y(i)}|H^{i},t_{ab}), & \text{if observations at positions}\\
 & \text{$x(i)$ and $y(i)$ in proteins $a$ and $b$,}\\
 & \text{respectively, are homologous;}\\
p(O_{a}^{x(i)}|H^{i}), & \text{if the observation at}
\\ & \text{position x(i) in protein \ensuremath{a}}\\
 & \text{is the result of an indel;}\\
p(O_{b}^{y(i)}|H^{i}), & \text{if the observation at} \\ & \text{position y(i) in protein \ensuremath{b}}\\
 & \text{is the result of an indel.}
\end{cases}
\end{align*}
$M_{ab}^{i}\in\left\{ \left(\substack{x\\
y
}
\right),\left(\substack{x\\
-
}
\right),\left(\substack{-\\
y
}
\right)\right\}$ specifies one of three possible homology relationships at position $i$ in the alignment (homologous amino acids, deletion with respect to protein $a$, and insertion with respect to protein $a$, respectively.). $x\in\{1,\ldots,|a|\}$ and $y\in\{1,\ldots,|b|\}$ specifies the indices of the positions in proteins $a$ and $b$, respectively.

\begin{figure}[h]
\iffigures
\centering
\includegraphics[height=0.3\textheight]{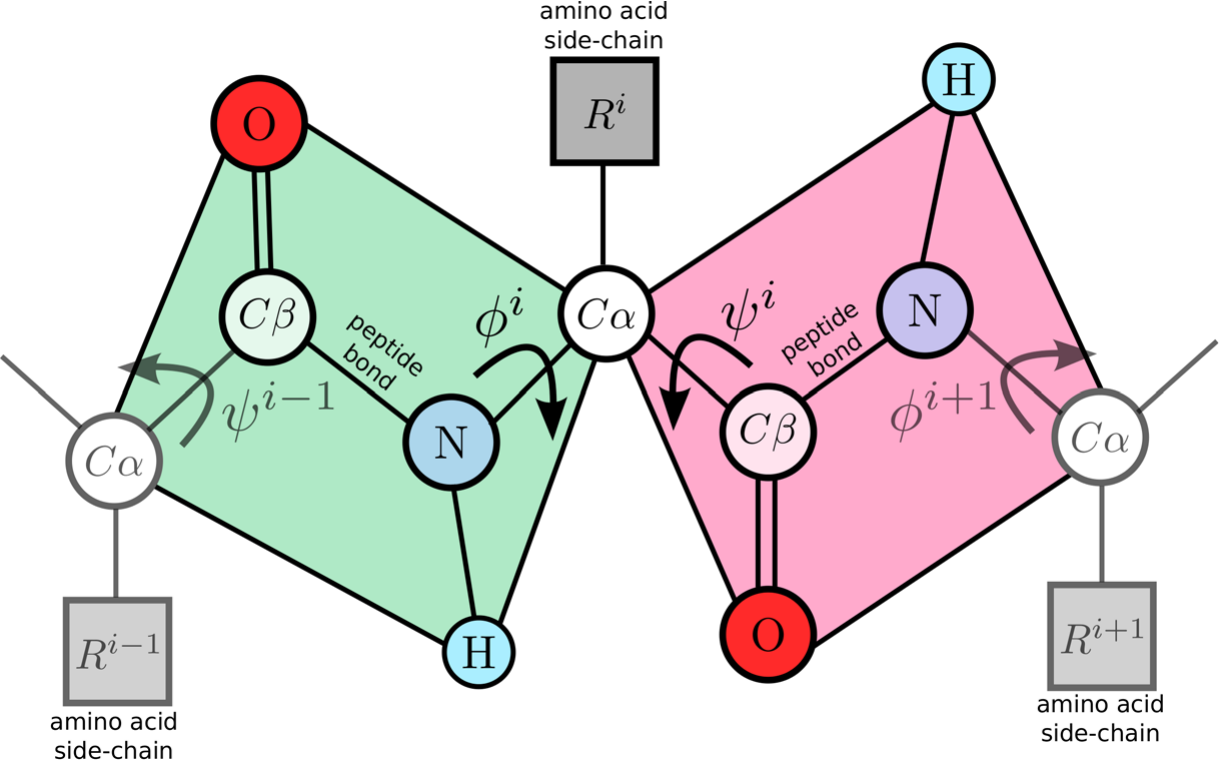}
\fi
\caption{\small Dihedral angle representation. A very small section of the protein backbone is displayed in atomic detail. The dihedral angles, $\phi$ and $\psi$, are shown. The partial  double bond nature of the peptide bond, gives rise to a planar configuration of O, C$\beta$, N and H atoms. This allows the protein backbone structure to be largely described in terms of a series of  $\phi$ and $\psi$ dihedral angles that relate the planes in three-dimensional space.}%
\label{fig:dihedralangles}%
\end{figure}

Note that a disadvantage of our approach is that neighbouring amino acid positions in the presence of a deletion or series of deletions  are no longer treated as directly adjacent by our HMM with respect to that particular protein, whereas in physical reality the amino acids positions would be directly adjacent to one another. However, the original HMM formulation where we treat the alignment as given \textit{a priori} has the same shortcoming.

\subsection{Parallelisation of model training}

The StEM algorithm is easily parallelised. In the E-step, the parameters, alignment and hidden states of protein pair proteins can be independently sampled in parallel, conditioned on the parameters, $\Psi^{(r)}$, from the M-step. Whereas, in the M-step the parameters corresponding to each hidden state can be independently updated in parallel conditioned on the samples from the E-step.

\subsection{Time-reversibility}

The three stochastic processes are assumed to be time-reversible. This, together with assumption of time-reversibility in jumping between evolutionary regimes in equation ensures overall time-reversibility. This allows to treat the phylogenetic tree relating proteins $p_{a}$ and $p_{b}$  as unrooted, implying we can arbitrarily pick $p_{a}$ or $p_{b}$  as a root of the phylogenetic tree \citep{felsenstein1981evolutionary}. This avoids the need to marginalise over the common ancestor protein of proteins $p_{a}$  and $p_{b}$. Note that whilst time-reversibility of the evolutionary processes at each site holds, this is different from reversibility of the HMM. The transition probability matrix of the HMM is not restricted to be reversible and therefore detailed-balance does not necessarily hold.

\begin{figure}[h]
\iffigures
\centering
\includegraphics[width=0.8\textwidth]{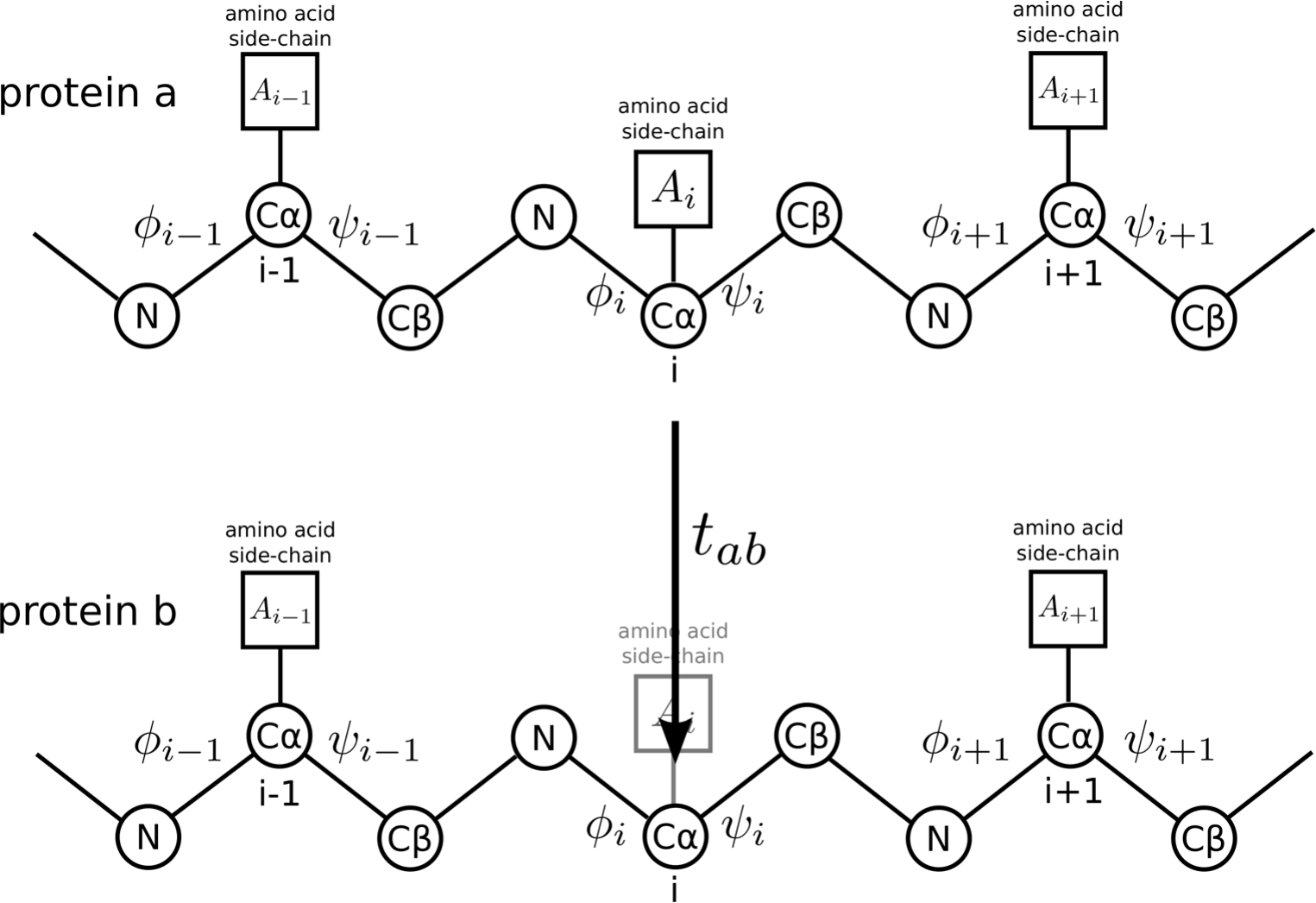}
\fi
\caption{\small Time-reversibility allows an arbitrary rooting of the phylogeny. The stochastic processes describing the observations have all been chosen such that time-reversibility holds. This is particularly useful in the case of a pairwise phylogeny, because it permits us to arbitrarily pick one of the extant proteins (protein $a$ or protein $b$) as the root of the phylogeny, without changing the likelihood of the data. This avoids computationally costly marginalisation of the unobserved common ancestor protein that in reality is shared by both of the extant proteins. Furthermore, only a single parameter, $t_{ab}$, the evolutionary time, need be marginalised when training the model or performing inference.}
\label{fig:timereversibility}
\end{figure}

\section{Supplementary results}

\subsection{Estimates of evolutionary time from dihedral angles are consistent with estimates from sequence}

Figure~\ref{fig:branch_length_whiskers} compares evolutionary times estimated using pairs of homologous amino acid sequences only versus pairs of homologous dihedral angles only. The mean 90\% confidence interval (CI) widths for the sampled evolutionary times were smaller for amino acid sequences (mean 90\% CI width $=0.139$) than dihedral angles (mean 90\% CI width $=0.191$).


\begin{thebibliography}{}

\bibitem[Arnold {\em et~al.}(2006)Arnold, Bordoli, Kopp, and
  Schwede]{arnold2006swiss}
Arnold, K., Bordoli, L., Kopp, J., and Schwede, T. (2006).
\newblock The {SWISS-MODEL} workspace: a web-based environment for protein
  structure homology modelling.
\newblock {\em Bioinformatics\/}, {22}(2):195--201.

\bibitem[Boomsma {\em et~al.}(2008)Boomsma, Mardia, Taylor, Ferkinghoff-Borg,
  Krogh, and Hamelryck]{boomsma2008generative}
Boomsma, W., Mardia, K.~V., Taylor, C.~C., Ferkinghoff-Borg, J., Krogh, A., and
  Hamelryck, T. (2008).
\newblock A generative, probabilistic model of local protein structure.
\newblock {\em Proceedings of the National Academy of Sciences\/}, {105}(26):8932--8937.

\bibitem[Challis and Schmidler(2012)Challis and
  Schmidler]{challis2012stochastic}
Challis, C.~J. and Schmidler, S.~C. (2012).
\newblock A stochastic evolutionary model for protein structure alignment and
  phylogeny.
\newblock {\em Molecular Biology and Evolution\/}, {29}(11):3575--3587.

\bibitem[Downs and Mardia(2002)Downs and Mardia]{downs2002circular}
Downs, T.~D. and Mardia, K. (2002).
\newblock Circular regression.
\newblock {\em Biometrika\/}, {89}(3):683--698.

\bibitem[Felsenstein(1985)Felsenstein]{felsenstein1985phylogenies}
Felsenstein, J. (1985).
\newblock Phylogenies and the comparative method.
\newblock {\em American Naturalist\/}, {125}(1):1--15.

\bibitem[Frellsen {\em et~al.}(2012)Frellsen, Mardia, Borg, Ferkinghoff-Borg,
  and Hamelryck]{frellsen2012towards}
Frellsen, J., Mardia, K.~V., Borg, M., Ferkinghoff-Borg, J., and Hamelryck, T.
  (2012).
\newblock Towards a general probabilistic model of protein structure: the
  reference ratio method.
\newblock In {\em Bayesian Methods in Structural Bioinformatics\/}, pages
  125--134. Springer.

\bibitem[Garc\'ia-Portugu\'es {\em et~al.}(2017)Garc\'ia-Portugu\'es,
  S\o{}rensen, Mardia, and Hamelryck]{garciap2017diffusions}
Garc\'ia-Portugu\'es, E., S\o{}rensen, M., Mardia, K.~V., and Hamelryck, T.
  (2017).
\newblock Langevin diffusions on the torus: estimation and applications.
\newblock {\em arXiv:1705.00296\/}.

\bibitem[Grishin(2001)Grishin]{grishin2001fold}
Grishin, N.~V. (2001).
\newblock Fold change in evolution of protein structures.
\newblock {\em Journal of Structural Biology\/}, {134}(2):167--185.

\bibitem[Hamelryck and Manderick(2003)Hamelryck and
  Manderick]{hamelryck2003pdb}
Hamelryck, T. and Manderick, B. (2003).
\newblock PDB file parser and structure class implemented in python.
\newblock {\em Bioinformatics\/}, {19}(17):2308--2310.

\bibitem[Herman {\em et~al.}(2014)Herman, Challis, Nov{\'a}k, Hein, and
  Schmidler]{herman2014simultaneous}
Herman, J.~L., Challis, C.~J., Nov{\'a}k, {\'A}., Hein, J., and Schmidler,
  S.~C. (2014).
\newblock Simultaneous Bayesian estimation of alignment and phylogeny under a
  joint model of protein sequence and structure.
\newblock {\em Molecular Biology and Evolution\/}, {31}(9):2251--2266.

\bibitem[Johnson(2014)Johnson]{johnson2014nlopt}
Johnson, S.~G. (2014).
\newblock The {NLopt} nonlinear-optimization package.

\bibitem[Mikl{\'o}s {\em et~al.}(2004)Mikl{\'o}s, Lunter, and
  Holmes]{miklos2004long}
Mikl{\'o}s, I., Lunter, G., and Holmes, I. (2004).
\newblock A ``long indel'' model for evolutionary sequence alignment.
\newblock {\em Molecular Biology and Evolution\/}, {21}(3):529--540.

\bibitem[Mizuguchi {\em et~al.}(1998)Mizuguchi, Deane, Blundell, and
  Overington]{mizuguchi1998homstrad}
Mizuguchi, K., Deane, C.~M., Blundell, T.~L., and Overington, J.~P. (1998).
\newblock HOMSTRAD: a database of protein structure alignments for homologous
  families.
\newblock {\em Protein Science\/}, {7}(11):2469--2471.

\bibitem[Nodelman {\em et~al.}(2002)Nodelman, Shelton, and
  Koller]{nodelman2002continuous}
Nodelman, U., Shelton, C.~R., and Koller, D. (2002).
\newblock Continuous time Bayesian networks.
\newblock In {\em Proceedings of the Eighteenth conference on uncertainty in
  artificial intelligence\/}, pages 378--387. Morgan Kaufmann Publishers.

\bibitem[Powell(1994)Powell]{powell1994cobyla}
Powell, M. J.~D. (1994).
\newblock A direct search optimization method that models the objective
  and constraint functions by linear interpolation. In \textit{Advances in Optimization and Numerical Analysis}, pages 51--67.
\newblock Springer.

\bibitem[Price {\em et~al.}(2010)Price, Dehal, and Arkin]{price2010fasttree}
Price, M.~N., Dehal, P.~S., and Arkin, A.~P. (2010).
\newblock {FastTree} 2--approximately maximum-likelihood trees for large
  alignments.
\newblock {\em PloS one\/}, {5}(3):e9490.

\bibitem[Rohl {\em et~al.}(2004)Rohl, Strauss, Misura, and
  Baker]{rohl2004protein}
Rohl, C.~A., Strauss, C.~E., Misura, K.~M., and Baker, D. (2004).
\newblock Protein structure prediction using {Rosetta}.
\newblock {\em Methods in Enzymology\/}, {383}:66--93.

\bibitem[Schwartz {\em et~al.}(2005)Schwartz, Myers, and
  Pachter]{schwartz2005alignment}
Schwartz, A.~S., Myers, E.~W., and Pachter, L. (2005).
\newblock Alignment metric accuracy.
\newblock {\em arXiv:q-bio/0510052\/}.

\bibitem[Thorne {\em et~al.}(1992)Thorne, Kishino, and
  Felsenstein]{thorne1992inching}
Thorne, J.~L., Kishino, H., and Felsenstein, J. (1992).
\newblock Inching toward reality: an improved likelihood model of sequence
  evolution.
\newblock {\em Journal of Molecular Evolution\/}, {34}(1):3--16.

\bibitem[Touw {\em et~al.}(2015)Touw, Baakman, Black, te~Beek, Krieger,
  Joosten, and Vriend]{touw2015series}
Touw, W.~G., Baakman, C., Black, J., te~Beek, T.~A., Krieger, E., Joosten,
  R.~P., and Vriend, G. (2015).
\newblock A series of {PDB-related} databanks for everyday needs.
\newblock {\em Nucleic Acids Research\/}, {43}(D1):D364--D368.

\bibitem[Whelan and Goldman(2001)Whelan and Goldman]{whelan2001general}
Whelan, S. and Goldman, N. (2001).
\newblock A general empirical model of protein evolution derived from multiple
  protein families using a maximum-likelihood approach.
\newblock {\em Molecular Biology and Evolution\/}, {18}(5):691--699.

\end{thebibliography}

\begin{thebibliography}{}

\bibitem[Edgar(2004)Edgar]{edgar2004muscle}
Edgar, R.~C. (2004).
\newblock {MUSCLE}: multiple sequence alignment with high accuracy and high
  throughput.
\newblock {\em Nucleic Acids Research\/}, {32}(5):1792--1797.

\bibitem[Felsenstein(1981)Felsenstein]{felsenstein1981evolutionary}
Felsenstein, J. (1981).
\newblock Evolutionary trees from {DNA} sequences: a maximum likelihood
  approach.
\newblock {\em Journal of Molecular Evolution\/}, {17}(6):368--376.

\bibitem[Katoh {\em et~al.}(2002)Katoh, Misawa, Kuma, and
  Miyata]{katoh2002mafft}
Katoh, K., Misawa, K., Kuma, K.-i., and Miyata, T. (2002).
\newblock MAFFT: a novel method for rapid multiple sequence alignment based on
  fast Fourier transform.
\newblock {\em Nucleic Acids Research\/}, {30}(14):3059--3066.

\bibitem[Mikl{\'o}s {\em et~al.}(2004)Mikl{\'o}s, Lunter, and
  Holmes]{miklos2004long}
Mikl{\'o}s, I., Lunter, G., and Holmes, I. (2004).
\newblock A ``long indel'' model for evolutionary sequence alignment.
\newblock {\em Molecular Biology and Evolution\/}, {21}(3):529--540.

\bibitem[Thorne {\em et~al.}(1992)Thorne, Kishino, and
  Felsenstein]{thorne1992inching}
Thorne, J.~L., Kishino, H., and Felsenstein, J. (1992).
\newblock Inching toward reality: an improved likelihood model of sequence
  evolution.
\newblock {\em Journal of Molecular Evolution\/}, {34}(1):3--16.

\end{thebibliography}

\begin{figure}[h]
\iffigures
\centering
\includegraphics[width=0.8\textwidth]{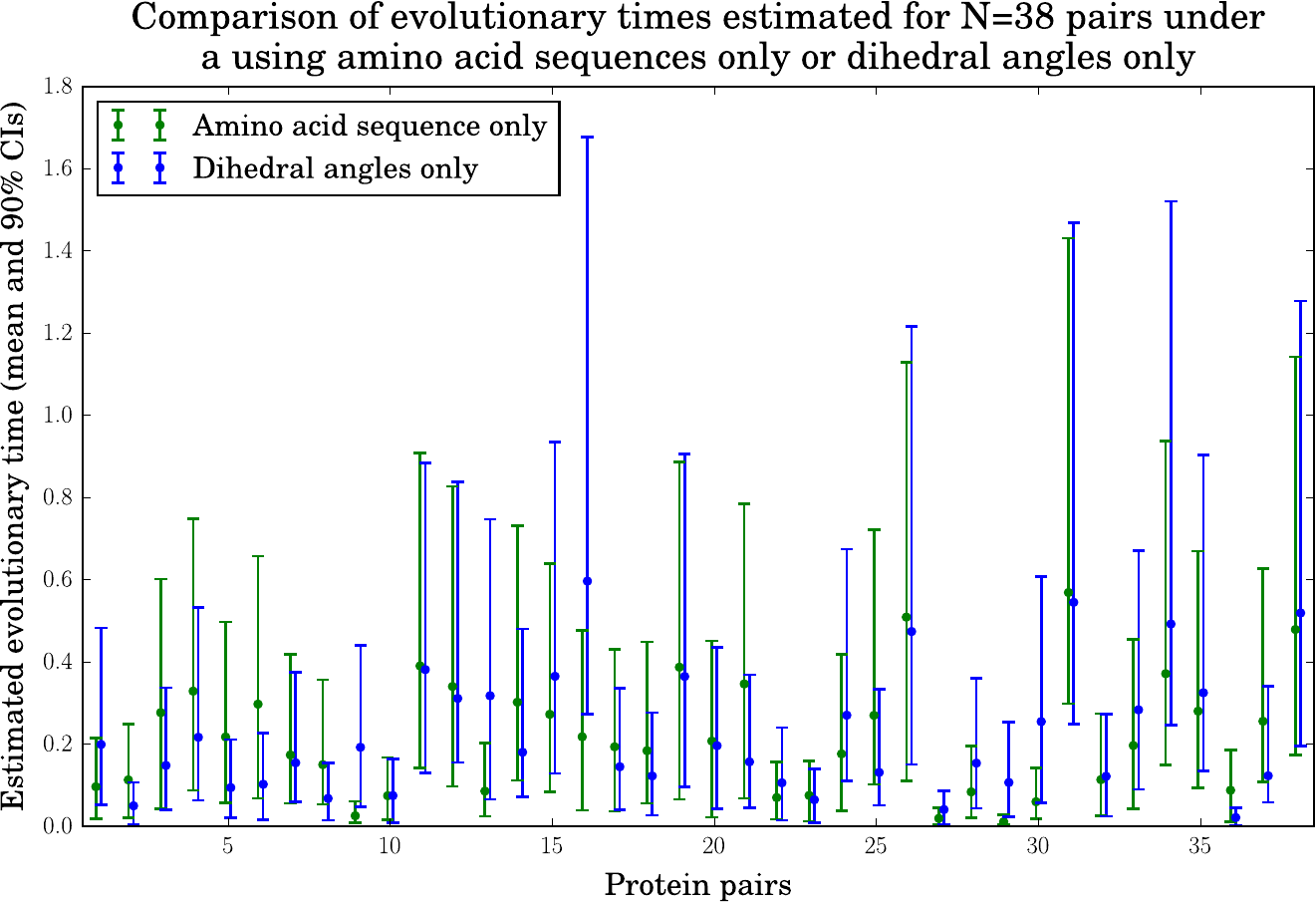}
\fi
\caption{\small Evolutionary times were estimated for 39 protein pairs under a model with jump evolution under two different conditions. Under the first condition where only the amino acid sequences were treat as observed, whereas under the second only the dihedral angles were treated as observed.}
\label{fig:branch_length_whiskers}%
\end{figure}

\fi
\end{document}